\documentclass[sigconf]{aamas}

\usepackage{balance} 
\usepackage{amsmath}
\usepackage{subcaption}
\usepackage{wrapfig}

\doi{UMCV8852}

\makeatletter
\gdef\@copyrightpermission{
  \begin{minipage}{0.2\columnwidth}
   \href{https://creativecommons.org/licenses/by/4.0/}{\includegraphics[width=0.90\textwidth]{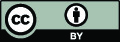}}
  \end{minipage}\hfill
  \begin{minipage}{0.8\columnwidth}
   \href{https://creativecommons.org/licenses/by/4.0/}{This work is licensed under a Creative Commons Attribution International 4.0 License.}
  \end{minipage}
  \vspace{5pt}
}
\makeatother

\setcopyright{ifaamas}
\acmConference[AAMAS '26]{Proc.\@ of the 25th International Conference
on Autonomous Agents and Multiagent Systems (AAMAS 2026)}{May 25 -- 29, 2026}
{Paphos, Cyprus}{C.~Amato, L.~Dennis, V.~Mascardi, J.~Thangarajah (eds.)}
\copyrightyear{2026}
\acmYear{2026}
\acmDOI{}
\acmPrice{}
\acmISBN{}

\title[AAMAS-2026 Formatting Instructions]{Theory of Mind Guided Strategy Adaptation for Zero-Shot Coordination}

\author{Andrew Ni}
\affiliation{
  \institution{Carnegie Mellon University}
  \city{Pittsburgh}
  \country{United States}}
\email{acni@andrew.cmu.edu}

\author{Simon Stepputtis}
\affiliation{
  \institution{Virginia Tech}
  \city{Blacksburg}
  \country{United States}}
\email{stepputtis@vt.edu}

\author{Stefanos Nikolaidis}
\affiliation{
  \institution{University of Southern California}
  \city{Los Angeles}
  \country{United States}}
\email{nikolaid@usc.edu}

\author{Michael Lewis}
\affiliation{
  \institution{University of Pittsburgh}
  \city{Pittsburgh}
  \country{United States}}
\email{mlewis@sis.pitt.edu}

\author{Katia P. Sycara}
\affiliation{
  \institution{Carnegie Mellon University}
  \city{Pittsburgh}
  \country{United States}}
\email{sycara@andrew.cmu.edu}

\author{Woojun Kim}
\affiliation{
  \institution{Carnegie Mellon University}
  \city{Pittsburgh}
  \country{United States}}
\email{woojunk@andrew.cmu.edu}

\begin{abstract}
A central challenge in multi-agent reinforcement learning is enabling agents to adapt to previously unseen teammates in a zero-shot fashion. Prior work in zero-shot coordination often follows a two-stage process, first generating a diverse training pool of partner agents, and then training a best-response agent to collaborate effectively with the entire training pool. While many previous works have achieved strong performance by devising better ways to diversify the partner agent pool, there has been less emphasis on how to leverage this pool to build an adaptive agent. One limitation is that the best-response agent may converge to a \textit{static, generalist} policy that performs reasonably well across diverse teammates, rather than learning a more \textit{adaptive, specialist} policy that can better adapt to teammates and achieve higher synergy. To address this, we propose an adaptive ensemble agent that uses Theory-of-Mind-based best-response selection to first infer its teammate's intentions and then select the most suitable policy from a policy ensemble. We conduct experiments in the Overcooked environment to evaluate zero-shot coordination performance under both fully and partially observable settings. The empirical results demonstrate the superiority of our method over a single best-response baseline.
\end{abstract}


\newcommand{\BibTeX}{\rm B\kern-.05em{\sc i\kern-.025em b}\kern-.08em\TeX}

\begin{document}


\pagestyle{fancy}
\fancyhead{}


\maketitle 


\section{Introduction}

Effective coordination with previously unseen partners is vital for multi‑agent systems deployed in real‑world settings. One promising approach is zero‑shot coordination (ZSC)~\cite{hu2020other, treutlein2021new, lupu2021trajectory}, in which agents must align with novel collaborators without any additional learning or explicit communication at test time. Achieving ZSC, however, is challenging because agents trained in self-play, where they are jointly trained, tend to overfit to shared conventions that emerge during learning. Consequently, they struggle to infer a new partner’s intent and often lack a principled strategy for adapting their behavior. This challenge is further exacerbated in deep learning–based approaches, as deep neural networks often behave unpredictably when exposed to unseen inputs~\cite{Eykholt_2018_CVPR,DBLP:journals/corr/GoodfellowSS14,pmlr-v97-engstrom19a,overcookedv2}, ultimately resulting in coordination failures. 

A common approach for ZSC is the population‑based approach~\cite{overcooked-ai,fcp,mep,gamma}, which first generates a diverse partner population, e.g., by using different seeds or by maximizing entropy across populations, and then trains a best-response (BR) agent\footnote{In game theory, the term best-response refers to the strategy which produces the most favorable outcome, when other players' strategies are fixed~\cite{fudenberg1991game,gibbons1992primer}} to maximize the expected return across the partner pool. During training, the BR agent interacts with partners sampled from the population, observes their behaviors, and learns to select actions that maximize the expected team reward averaged across all possible partners. In other words, the BR agent implicitly infers each partner’s policy through experience and adapts its own actions to achieve the highest overall return when paired with diverse partners. Despite reasonable performance in many tasks, this approach still has a limitation: the BR agent is trained to fit all partners with a single policy and may converge to a more \textit{static, generalist} policy that does not fully adapt to the new partners. Intuitively, the static generalist policy represents an easy-to-learn local optimum that hinders the BR agent from learning a truly adaptive policy. Effective collaboration, however, inherently requires adapting to the behaviors of one’s partners, rather than relying on fixed conventions or averaged responses. The ability to adjust one’s policy based on the partner’s behavior is essential for achieving robust coordination with diverse or unseen partners. Our goal is to develop a method that enables an agent to accurately infer a partner’s intent and adapt its actions in a zero-shot manner.

In this paper, we propose the \textbf{T}heory of Mind-based \textbf{B}est Response \textbf{S}election (\textbf{TBS}) framework 
to address the adaptivity gap in population-based BR methods. 
TBS leverages the concept of Theory of Mind (ToM)---the ability to infer others’ beliefs and intentions---to enable agents to reason about their partners’ behavior and adapt their own actions accordingly. This capability enhances adaptivity and leads to more effective coordination with novel or dynamically changing partners. Building on this insight, TBS constructs a diverse partner pool, clusters partners into behaviorally distinct groups through self-tuning spectral clustering, and trains a specialized BR policy for each cluster. At execution time, a ToM-based adaptive selection module infers a partner’s intent and identifies the most similar cluster, selecting the corresponding BR policy for real-time coordination. By integrating ToM reasoning with population-based training, TBS bridges the adaptivity gap and achieves robust zero-shot coordination with previously unseen partners.

We evaluate TBS in terms of ZSC performance in the Overcooked environment~\cite{overcooked-ai}, a fully cooperative two-player cooking game. In addition, we conduct ablation studies assessing the performance of our method with different training population sizes and with different hyperparameter choices such as different number of clusters. Empirical results show that the proposed method outperforms the population-based Best-Response method~\cite{mep,fcp}, with a performance gap that increases with the size of the training pool. Overall, our results demonstrate the utility of explicit ToM based adaptation for zero-shot coordination. Our contributions include: \\
$\bullet$ Automatic identification of strategy clusters within a diverse partner population using self-tuning spectral clustering of cross-play performance.\\
$\bullet$ A Theory-of-Mind-based adaptive selection module that infers a partner’s intent and dynamically selects the most suitable best-response policy from an ensemble.\\
$\bullet$ Demonstration of TBS’s adaptivity and generalization, showing that combining strategy-specific BR agents with ToM-guided policy selection yields robust ZSC in the Overcooked environment.

\section{Background and Related Works}

\subsection{Multi-Agent Reinforcement Learning}

A decentralized partially observable Markov decision process (Dec-POMDP)~\cite{decpomdp,intro-decpomdp} formalizes fully cooperative multi-agent decision-making under partial observability. At each timestep $t$, each agent $i$ receives a local observation $o_t^i$ of the global state $s_t$ and selects an action according to its policy $a_t^i \sim \pi^i(\cdot \mid o_t^i)$. The environment then transitions to a new state $s_{t+1}$ and all agents receive a shared reward $r_t$ that depends on the joint action. The objective is to maximize the expected discounted return under the joint policy $(\pi^1, \cdots, \pi^N)$:
\begin{align}\label{eq:self-play}
    J(\pi^1, \cdots, \pi^N) = \mathbb{E}_{\tau \sim (\pi^1, \cdots, \pi^N)}\Big[ \sum_{t=0}^{\infty} \gamma^t r_t \Big],\vspace{-2ex}
\end{align}
where $\gamma$ and $\tau$ denote the discount factor and a trajectory sampled from the trajectory distribution induced by the joint policy $(\pi^1, \ldots, \pi^N)$ interacting with the environment dynamics, respectively. In standard cooperative MARL, these policies are typically trained jointly in \textit{self-play}, where all agents learn simultaneously within the same environment instance.

A common framework for Dec-POMDPs is centralized training with decentralized execution (CTDE), where agents are trained with access to global information but execute policies using only local observations~\cite{jeon2022maser, kim2023adaptive}. Under CTDE, a key research direction in cooperative MARL focuses on enhancing coordination among agents, such as maximizing mutual information~\cite{li2022pmic, kim2023variational} learning to communicate~\cite{jiang2018learning, kim2019message}, 
and value decomposition methods~\cite{vdn, qmix}. VDN~\cite{vdn} is a value decomposition method for multi-agent credit assignment that approximates the joint Q-function as a sum of individual value functions, and trains it by minimizing the temporal-difference error.

\subsection{Zero-Shot Coordination}

Despite the success of various approaches in cooperative multi-agent environments,
the learned policies may fail to cooperate even with other agents trained using the same algorithm but different random seeds. This limitation motivates the problem of zero-shot coordination~\cite{other-play,overcooked-ai}, where agents should be able to cooperate with novel partners without additional training. This stems from the fact that, as described in Eq.~\ref{eq:self-play}, standard cooperative MARL optimizes the expected return $J(\pi^1, \cdots, \pi^N)$ through self-play, where all agents are trained jointly and their policies co-adapt to one another.
Consequently, the trajectory distribution $\tau \sim (\pi^1, \cdots, \pi^N)$ used during training is determined by these co-adapted policies. When an agent is later paired with an independently trained partner $\hat{\pi}^j$, the induced trajectory distribution $\tau' \sim (\pi^1, \cdots, \hat{\pi}^j, \ldots, \pi^N)$ differs from the one optimized during training.
As a result, the optimization objective no longer matches the evaluation setting, leading to a shift in the trajectory distribution and potential coordination failure.

For evaluation, ZSC performance is commonly measured using the cross-play (XP) and inter-algorithm cross-play (inter-XP) objectives.
In the two-agent case, the XP objective assesses how well agents trained independently under the same learning algorithm $\mathcal{A}$ can coordinate:
\begin{align}
    J_{\text{XP}}(\mathcal{A})=\mathbb{E}_{(\pi^1_A,\pi^2_A),(\pi^1_B,\pi^2_B)\overset{\mathrm{iid}}{\sim} \mathcal{A}}\left[\frac{J(\pi^1_A,\pi^2_B)+J(\pi^1_B,\pi^2_A)}{2}\right].
\end{align}
Here, $(\pi^1_A, \pi^2_A)$ and $(\pi^1_B, \pi^2_B)$ denote two independent policy pairs obtained from separate training runs of the same algorithm~$\mathcal{A}$ with different random seeds.
The expectation thus measures how well independently trained agents from~$\mathcal{A}$ can coordinate when paired across runs.

Prior works such as Other-Play~\cite{other-play} and Off-Belief Learning~\cite{obl} optimize their methods to improve performance under this metric.
More recently, Any-Play~\cite{any-play} extends the evaluation to partners trained with different algorithms, defining the inter-XP objective as $J_{\text{inter-XP}}(\mathcal{A})=$
\begin{align}
\mathbb{E}_{(\pi^1_A,\pi^2_A)\sim\mathcal{A},(\pi^1_B,\pi^2_B)\sim\mathcal{B} }\left[\frac{J(\pi^1_A,\pi^2_B)+J(\pi^1_B,\pi^2_A)}{2}\right]
\end{align}
where $\mathcal{A}$ and $\mathcal{B}$ denote two distinct multi-agent learning algorithms, each providing independently trained policy pairs. To improve ZSC performance, a variety of approaches have been proposed, including convention-avoidance methods, which aim to reduce agents’ reliance on arbitrary conventions formed during self-play, and population-based methods~\cite{overcooked-ai,fcp,mep,gamma}, which train cooperative agents against a diverse set of partners to promote generalization and adaptability. We next discuss population-based methods in more detail.

\begin{figure*}[t]
  \centering
  \begin{subfigure}[m]{0.5\textwidth}
        \centering
        \includegraphics[width=0.75\textwidth]{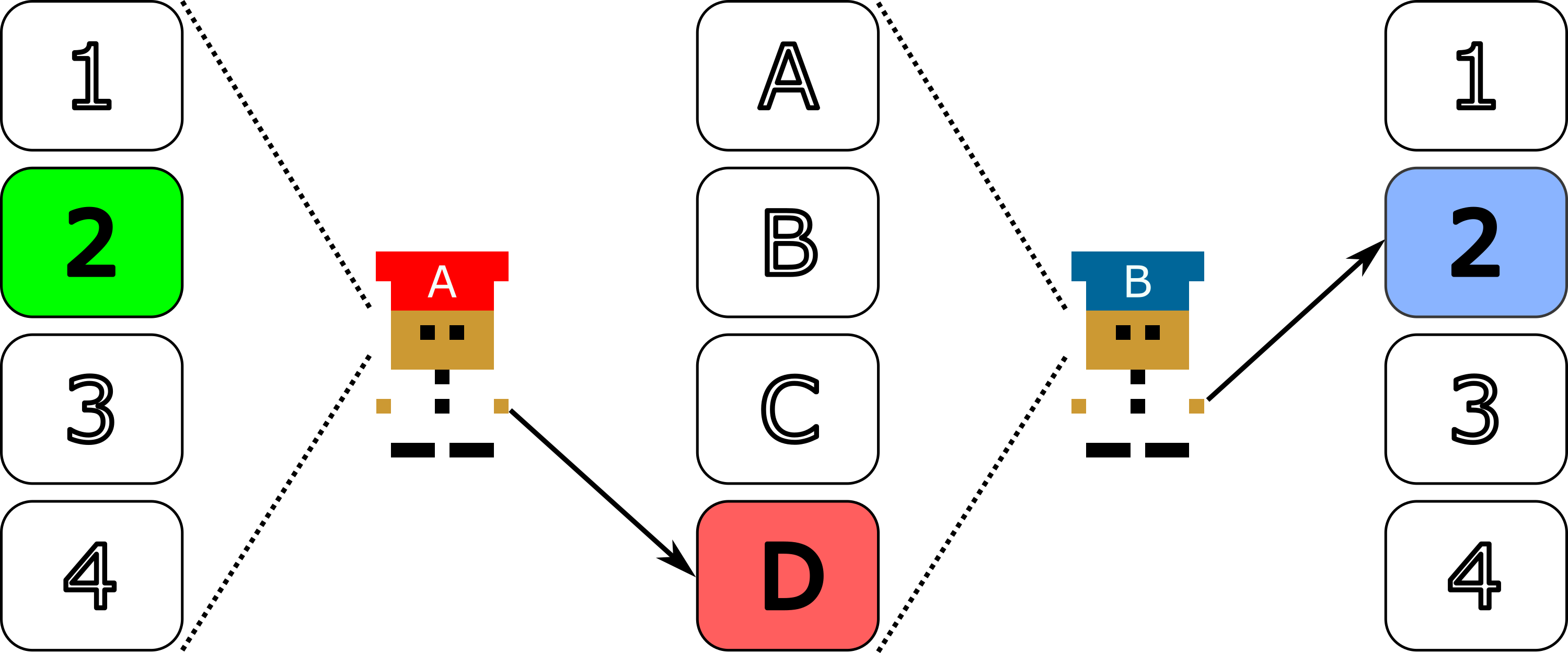}
        \vspace{3ex}
        \caption{Toy Communication Environment}
    \end{subfigure}%
    \hfill
\begin{subfigure}[m]{0.5\textwidth}
        \centering
\includegraphics[width=0.55\textwidth]{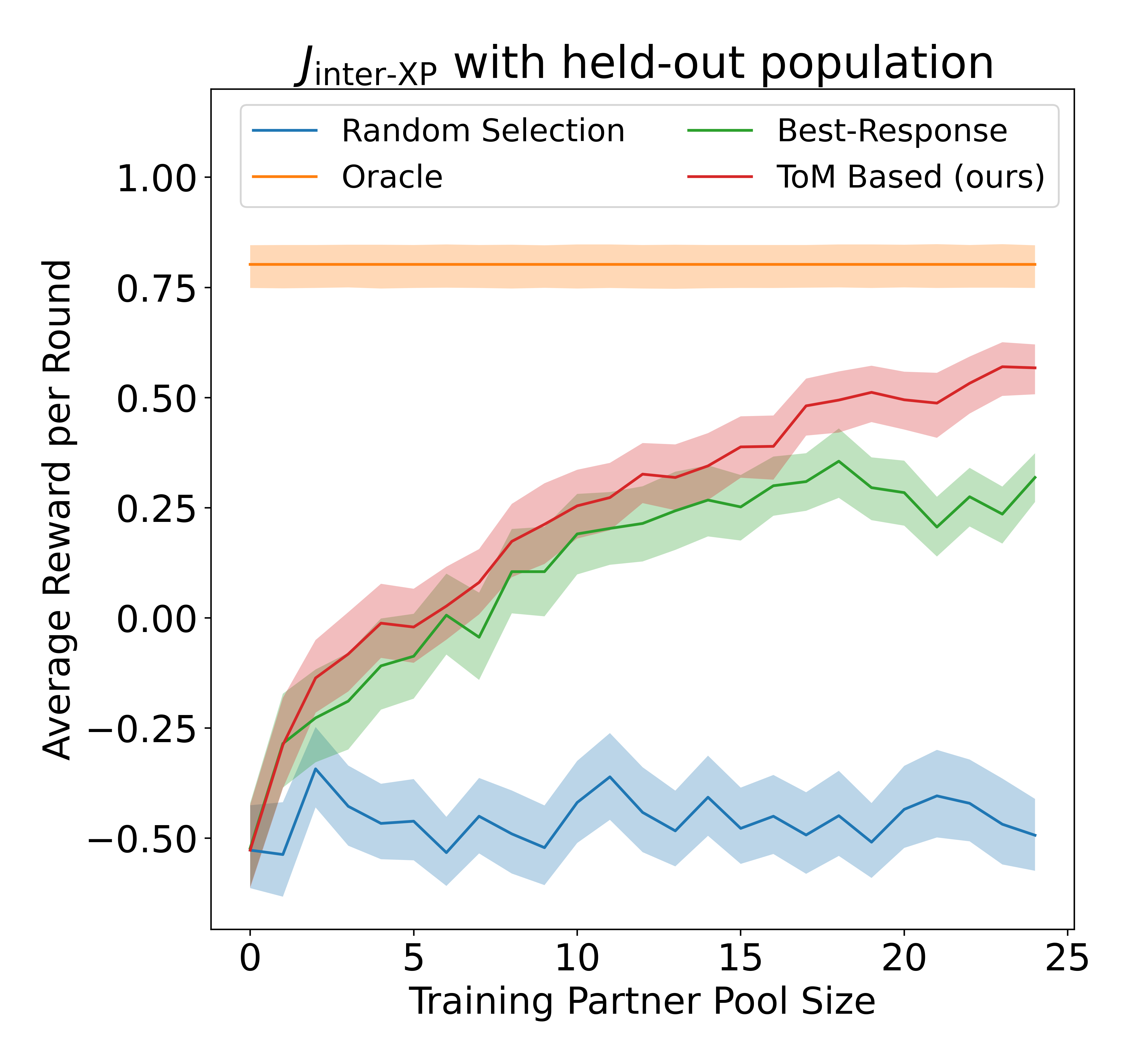}
        \caption{$J_{\text{inter-XP}}$ ZSC Performance}
    \end{subfigure}
  
  \caption{{\bf Left:} Toy communication environment. Alice (in red) is shown a random number and can take actions A through D or bail out. Bob (in blue) sees Alice's action and can guess the random number or bail out. After the round is over, the random number is revealed. {\bf Right:} $J_{\text{inter-XP}}$ of different PBT ZSC algorithms as a function of partner agent pool size in a toy environment. Shaded regions represent bootstrapped 95\% confidence intervals capturing variability over evaluation seeds.}
  \Description{Figure 1. Fully described in text}
  \label{fig:1}
\end{figure*}

\textbf{Population-based ZSC}~~has been widely investigated as a framework for enabling adaptive agents to achieve ZSC with novel partners at test time~\cite{pbt, li2025modeling}. The PBT approach typically involves (1) creating a \textit{diverse population} of simulated agents, and (2) training an adaptive cooperator agent as a \textit{best response} to this population to mitigate overfitting to specific strategies and promote generalization across diverse partners. 
Methods for constructing a diverse partner population~\cite{fcp, mep, trajedi,celebrating-diversity} include Fictitious Co-Play~\cite{fcp}, which trains multiple agent pairs with different seeds and takes checkpoints from the start, middle, and end of training to represent partners with diverse skill levels, Maximum Entropy Population-based (MEP) training~\cite{mep}, which maximizes the entropy of the average policy of the population, random reward shaping~\cite{hsp,zsceval,tsf}, which models implicit biases by adding a random reward shaping to each agent during self-play training, and other diversity-promoting techniques such as 
training a generative model to simulate infinitely many partner agents~\cite{gamma}. In this work, we use random reward shaping to generate a diverse population of agents.

Once the population is constructed, most population-based approaches train a single cooperator,i.e., BR agent, to learn a common best response to the entire population of simulated partners~\cite{mep,fcp,gamma,overcooked-ai}.
However, this design may limit the BR agent’s ability to adapt to novel partners, as it tends to learn a static generalist policy rather than an adaptive one~\cite{pecan,strategy-matching}.
To improve adaptability, \citet{pecan} and \citet{learning-representations} proposed conditioning the BR policy on a representation of the partner derived from prior interactions with that partner.
Specifically, PECAN~\cite{pecan} learns a few-shot, level-based best response conditioned on the approximate skill level of a given partner, which is estimated by a context encoder using trajectory data from previous episodes.
In contrast, PALM~\cite{learning-representations} introduces a framework that adapts to unknown partners by conditioning the BR policy on a latent embedding of a partner’s policy, which is fine-tuned at test time using data from prior interactions. Similarly, TSF~\cite{tsf} and MESH~\cite{strategy-matching} construct a library of diverse strategies and adapt to unknown partners within a single episode by evaluating how well each strategy explains the partner’s recent actions.
TSF builds its strategy library through a mix of heuristic agents and RL with reward shaping, selecting an appropriate strategy to execute depending on partner strategy, whereas MESH uses apprenticeship learning from clustered human–human trajectories and performs a weighted combination of strategies during execution.

\subsection{Theory of Mind}\label{sec:tom}

Theory of Mind~\cite{tom} refers to the ability of an agent to reason about the mental states of itself and others, including their intentions, beliefs, and preferences.
Recent work has shown that such capabilities can be approximated by deep neural networks.
For example, Theory of Mind Networks (ToMNets)~\cite{mtom} model other agents solely from observed behavior, enabling a learned prior over plausible behavioral patterns.
These networks can improve their predictive accuracy after observing a few timesteps of another agent’s behavior and infer false beliefs in partially observable scenarios.

Forming such internal models of other agents, considered as agent modeling, is crucial for adaptive multi-agent interaction~\cite{opponent-modelling-survey,kim2021communication}.
In this work, we focus on a form of ToM that infers an agent’s intentions, represented as a set of interpretable concepts ${c[i]}_{i=1}^{M}$, where each $c[i]$ corresponds to one of $M$ predefined, environment-specific intentions~\cite{cbm,concept-based-explanations,tcav}.
For instance, in the Overcooked environment, a concept may represent the “intention to pick up an onion from the onion pile.” The ToM network outputs the probability of each concept being active given an observation, and is trained using binary cross-entropy loss against the corresponding ground-truth concept labels obtained from the environment.
The loss function for the ToM parameters $\theta_{\text{ToM}}$ is given by
\begin{align}\label{eq:tom_train}
\mathcal{L}(\theta_{\text{ToM}}) =
&\mathbb{E}_{(o_t, c_t)}
\Big[c_t \log \hat{c}_t + (1 - c_t) \log (1 - \hat{c}_t) ,\Big]  \\ &\text{where} \quad
\hat{c}_t = \text{ToM}(o_t; \theta_{\text{ToM}}). \nonumber
\end{align}
This objective encourages the ToMNet to accurately predict the partner’s high-level intentions from observations

\begin{figure*}[t]
  \centering
  \includegraphics[width=0.95\textwidth]{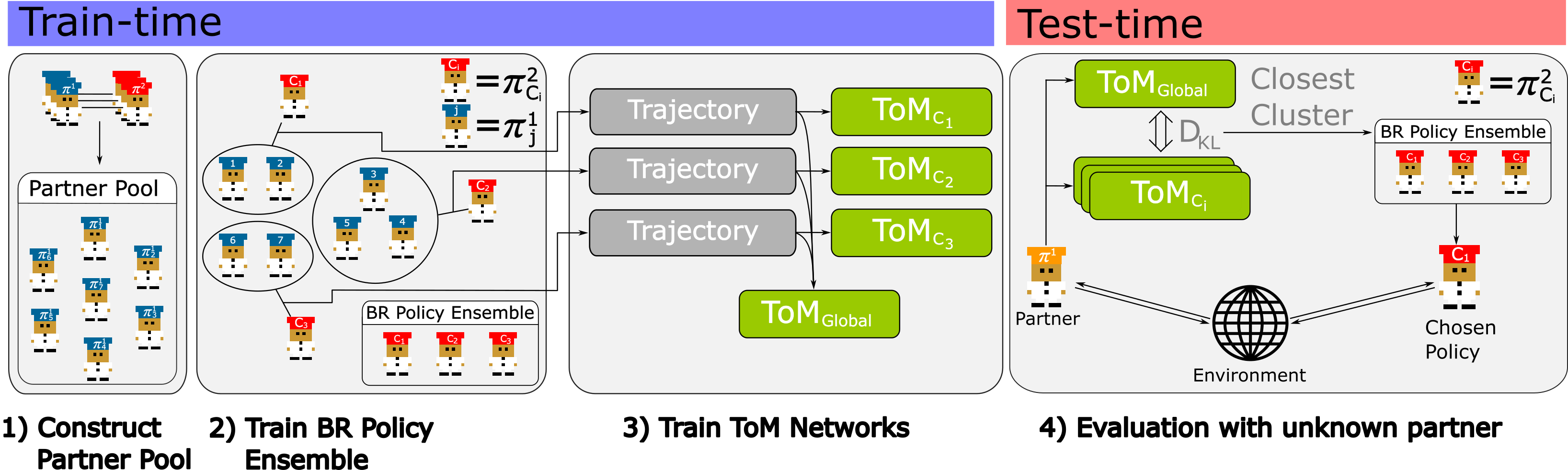}
  \caption{Overview of TBS: During training, (1) a diverse partner pool is constructed, (2) partners are clustered by behavioral similarity and specialist BR policies are trained per cluster, and (3) global and cluster-specific ToM networks are learned. During testing, the cooperator infers an unseen partner’s cluster and selects the corresponding BR policy for adaptive coordination.}
  \Description{Figure 2. Fully described in text}
  \label{fig:3}
\end{figure*}

\section{Methodology}

\subsection{Motivation}\label{sec:motivation}
One limitation of training a BR against an entire population of partners is that it often converges to a static, generalist policy rather than an adaptive one~\cite{pecan,strategy-matching}.
Because adaptive behavior requires learning multiple distinct strategies and maintaining long-term memory for partner modeling, such generalist policies can easily become local optima during BR training.
Moreover, as the diversity and size of the partner pool increase, learning an adaptive BR becomes increasingly difficult, diminishing the potential ZSC benefits of population diversity.


To illustrate these limitations, we consider a simple two-agent communication environment (Fig.~\ref{fig:1} (a)) with 16 rounds.
In each round, the red agent (Alice) is shown a random number from 1 to 4 and can either take one of four actions (A-D) or bail out for a reward of 0. The blue agent (Bob) observes Alice's action and must either guess Alice's number, receiving a reward of 1 if correct and -1 if incorrect, or bail out for a reward of 0; after each round, the true number is revealed to Bob. This task can be viewed as a signaling game: Alice's action (A-D) serves as a discrete message intended to convey the observed number (1-4). Each self-played pair develops its own private convention mapping numbers to actions (e.g., Alice 1$\to$A, 2$\to$B, 3$\to$C, 4$\to$D with Bob decoding A$\to$1, B$\to$2, C$\to$3, D$\to$4), while a different pair may learn a different convention (e.g., 1$\to$A, 2$\to$C, 3$\to$B, 4$\to$D). Consequently, ZSC across pairs is non-trivial: Bob must infer the partner’s convention online after observing Alice’s signal.

In the toy environment, we consider the task of adapting as Bob to an unseen Alice agent.
We use MEP~\cite{mep} to generate a training pool of Alice agents and a separate held-out pool for evaluation.
We compare four methods:
(i) \textit{Random Selection}, which selects a random strategy from the pool;
(ii) \textit{BR agent}, which trains a single agent to cooperate with any agent in the Alice pool;
(iii) \textit{ToM-based Best-Response Selection (TBS, ours)}, which infers Alice’s intentions via ToM and selects the specialized Bob agent that best matches them; and
(iv) \textit{Oracle}, which uses the exact partner co-trained with the teammate, providing an upper bound on ZSC performance.
Fig.~\ref{fig:1} (b) shows the inter-XP performance $J_{\text{inter-XP}}$ of the four methods as a function of the number of training agents.
When the pool is small, our method performs comparably to the BR baseline.
However, as the pool grows, the BR agent struggles to adapt to the increasing diversity of strategies and instead converges to a static generalist policy.
In contrast, our approach, through ToM-based reasoning, retains strong adaptation capability and outperforms both the BR and random baselines.

\subsection{ToM-based Best Response Selection}\label{sec:tbs}

To overcome the limitations of static generalist BR policy, 
we propose an adaptive \textbf{T}heory of Mind-based \textbf{B}est Response \textbf{S}election (\textbf{TBS}) framework. 
TBS trains an ensemble of specialist BR policies and adaptively selects one to act based on the inferred intent of a new teammate through ToM reasoning. 
To this end, TBS includes a cooperator agent that collaborates with partners at test time. 
It comprises three main components: (1) a diverse partner pool, (2) an ensemble of specialist BR policies, and (3) an adaptive selection module that dynamically selects the most suitable policy 
from the ensemble according to the observed behavior of the teammate. Fig.~\ref{fig:3} provides an overview of the proposed TBS framework.

\textbf{(1) Partner Pool Construction:} ~We begin by constructing a diverse pool of simulated partner agents using population-based training methods such as Fictitious Co-Play (FCP)~\cite{fcp}, Maximum Entropy Population-based training~\cite{mep}, and random reward shaping~\cite{reward-randomization,hsp,zsceval}, following prior BR approaches. 
Rather than training a single generalist BR policy against this population, 
we train an ensemble of BR policies, each specialized to cooperate with a distinct strategy cluster within the partner pool.

\begin{figure*}[t]
  \centering
  \includegraphics[width=\textwidth]{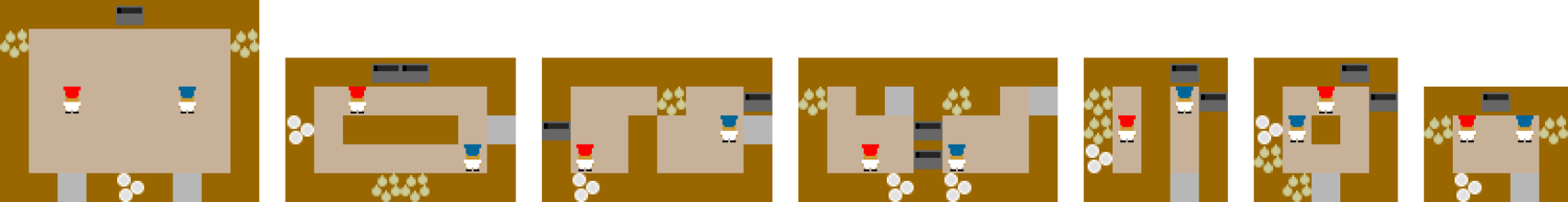}
  \caption{Onion soup environment layouts. From left to right: \textit{Large Room}, \textit{Counter Circuit}, \textit{Bothway Coordination}, \textit{Asymmetric Advantages}, \textit{Forced Coordination}, \textit{Coordination Ring}, \textit{Cramped Room}.}
  \Description{Figure 3. The Large room layout is a square layout with a pot on the top, onions on two sides, and plates and serving counters at the bottom. The counter circuit is oval ring shaped layout with pots at the top, plates on the left, onions at the bottom, and a serving counter on the right. Bothway coordination, Asymmetric Advantages, and Forced coordination all feature separations of the agents into two different spaces. Coordination ring features a ring shaped layout with a counter in the middle, and cramped room is a small 2x3 room.}
  \label{fig:layouts}
\end{figure*}

\textbf{(2) BR Policy Ensemble:} 
Having constructed a diverse partner pool, 
we now train an ensemble of BR policies, 
each specialized to collaborate with behaviorally similar groups of partners. 
Rather than training a separate BR policy for every individual partner, 
we group partners into clusters to improve computational efficiency 
and promote generalization across similar strategies. For this, we first cluster the partner pool into distinct strategy clusters using 
\textit{self-tuning spectral clustering}~\cite{stsc}, 
which automatically determines the number of clusters based on the policy similarity matrix. 
Specifically, we construct the similarity matrix using cross-play performance, 
where each entry is defined as
\begin{equation}
    \label{eq:cross-play-similarity}
    s(A,B) = \frac{J(\pi_A^1, \pi_B^2) + J(\pi_B^1, \pi_A^2)}
    {J(\pi_A^1, \pi_A^2) + J(\pi_B^1, \pi_B^2)}.
\end{equation}
Intuitively, the better two agent pairs $A$ and $B$ perform in cross-play, the more similar they are, 
and thus the closer they are in the similarity space. 
Applying self-tuning spectral clustering (STSC)~\cite{stsc} to the cross-play similarity matrix yields $k$ clusters, 
denoted by $\mathcal{C}_1, \ldots, \mathcal{C}_k$, 
where each cluster contains a subset of agent pairs with similar behavior. Specifically, given an $n \times n$ similarity matrix $S$, where $s(A,B)$ denotes each entry, 
STSC~\cite{stsc} performs eigen-decomposition 
of the graph Laplacian of $S$ and collects the top $k$ eigenvectors $x_1, \ldots, x_k$ to form the matrix $X = [x_1, \ldots, x_k] \in \mathbb{R}^{n \times k}$. In the ideal case where the Laplacian is strictly block-diagonal, 
each row of $X$ has only one nonzero entry (up to some rotation $XR$, with $R \in \mathbb{R}^{k \times k}$). STSC defines the alignment cost for a given number of clusters $k$ and rotation matrix $R$ as
\begin{equation}
    J(R, k) = \sum_{i=1}^{n} \sum_{j=1}^{k} 
    \frac{(XR)_{ij}^2}{[\max_j (XR)_{ij}]^2},
    \label{eq:stsc}
\end{equation}
and determines the optimal number of clusters 
by first minimizing this cost over all rotation matrices $R$ 
for each $k$, and then selecting the $k$ that yields the lowest alignment cost. Here, $k$ corresponds to the number of strategy clusters used in our BR ensemble.

We then train a BR policy for each cluster. 
Specifically, for the BR policy corresponding to cluster~$i$, with parameters $\theta^{i}_{\mathrm{BR}}$, 
we optimize its policy $\pi_{\theta^{i}_{\mathrm{BR}}}$ to maximize the expected return when paired with partners from cluster~$i$. 
The training objective is defined as
\begin{align}
    J_{\mathrm{BR}}(\theta^{i}_{\mathrm{BR}}) =
    \mathbb{E}_{A \sim U(\mathcal{C}_i),~\tau \sim [\pi_A^1, \pi_{\theta^{i}_{\mathrm{BR}}}^2]}
    \Big[\sum_{t=0}^{\infty} \gamma^t r_t\Big],
\end{align}
where $U(\mathcal{C}_i)$ denotes the uniform distribution over the agent pairs in cluster~$\mathcal{C}_i$. 
Note that the BR policies are complementary to their corresponding clusters.

{\bf (3) Adaptive Selection:} 
For adaptation, the cooperator agent should be able to infer the partner’s intent and act accordingly. 
In the adaptive selection module, we use ToM models to infer the partner’s intent 
and determine the most behaviorally similar cluster, allowing the agent to select and execute 
the corresponding complementary BR policy.

To do this, we build a total of $k+1$ ToM models: $k$ cluster-specific ToM models, denoted as $\text{ToM}_{C_i}$ for each cluster~$i$, 
which infer the partner’s intent within their respective clusters, 
and one global ToM model, denoted as $\text{ToM}_{\mathrm{global}}$, 
which infers partners across all clusters. Here, all ToM models are implemented as recurrent networks. In addition, 
each $\text{ToM}_{C_i}$ is trained to output a concept representation (i.e., a high-level action) 
corresponding to the intention of a partner agent within cluster~$i$, 
whereas $\text{ToM}_{\mathrm{global}}$ is trained to predict a concept representation of 
the partner’s most likely intent, inferred from the history and generalized across clusters. To train the ToM models, we generate trajectories from each cluster and use them to train $\text{ToM}_{C_i}$ with its respective trajectories, 
and $\text{ToM}_{\mathrm{global}}$ with trajectories aggregated across all clusters. All ToM models are optimized according to Eq.~\ref{eq:tom_train}, as described in Sec.~\ref{sec:tom}.

The outputs of the $k{+}1$ ToM models are leveraged at test time 
in the following online adaptation phase to infer which cluster the new partner corresponds to.

\subsection{Online Adaptation to New Partners}\label{sec:onlineadaptation}

The rationale behind TBS’s adaptation to new partners is that effective coordination requires understanding the partner’s intent and aligning one’s behavior accordingly. 
By leveraging ToM reasoning to infer the partner’s underlying intention and identify its corresponding cluster, 
TBS enables the agent to select an appropriate BR policy that matches the partner’s behavior. Specifically, the corresponding cluster is identified by comparing the output of the global ToM model 
with those of the cluster-specific ToM models; 
the cluster whose ToM output is closest to that of the global model is regarded as the one corresponding to the new partner. Formally, we compute the divergence between the Bernoulli distributions of the concept predictions from the cluster-specific and global ToM models as $D(\text{ToM}_{C_i}, \text{ToM}_{\mathrm{global}}) =$
\begin{equation}\label{eq:onlineadapt}
    \sum_{t'=1}^{t} \sum_{j=1}^{M} 
    D_{\mathrm{KL}} \big( 
        \text{Bernoulli}(c^i_{t'}[j]) 
        \;\|\; 
        \text{Bernoulli}(c^{global}_{t'}[j]) 
    \big),
\end{equation}
where $c^i_{t'}$ and $c^{global}_{t'}$ denote the concept distributions predicted by $\text{ToM}_{C_i}$ and $\text{ToM}_{\mathrm{global}}$, respectively, and $D_{\mathrm{KL}}(\cdot|\cdot)$ denotes the Kullback–Leibler divergence.

We then select the index $\hat{i}$ corresponding to the minimum distance, 
$\hat{i} = \arg\min_i D(\text{ToM}_{C_i}, \text{ToM}_{\mathrm{global}})$, 
and assign the $\hat{i}$-th BR policy to collaborate with the new partner. 
Since the BR policies in our ensemble are specialized for distinct strategy clusters identified by the clustering algorithm, 
this process can be intuitively viewed as identifying the teammate’s strategy and choosing the most appropriate response. 
At the beginning of each episode, the cooperator agent randomly selects one policy from the ensemble, 
as some interaction history is required before the partner’s intent can be inferred.

\begin{figure*}[t!]
    \centering
    \hspace{-5ex}
    \begin{subfigure}[b]{0.48\textwidth}\vskip 0pt
        \centering
        \includegraphics[width=1.\textwidth]{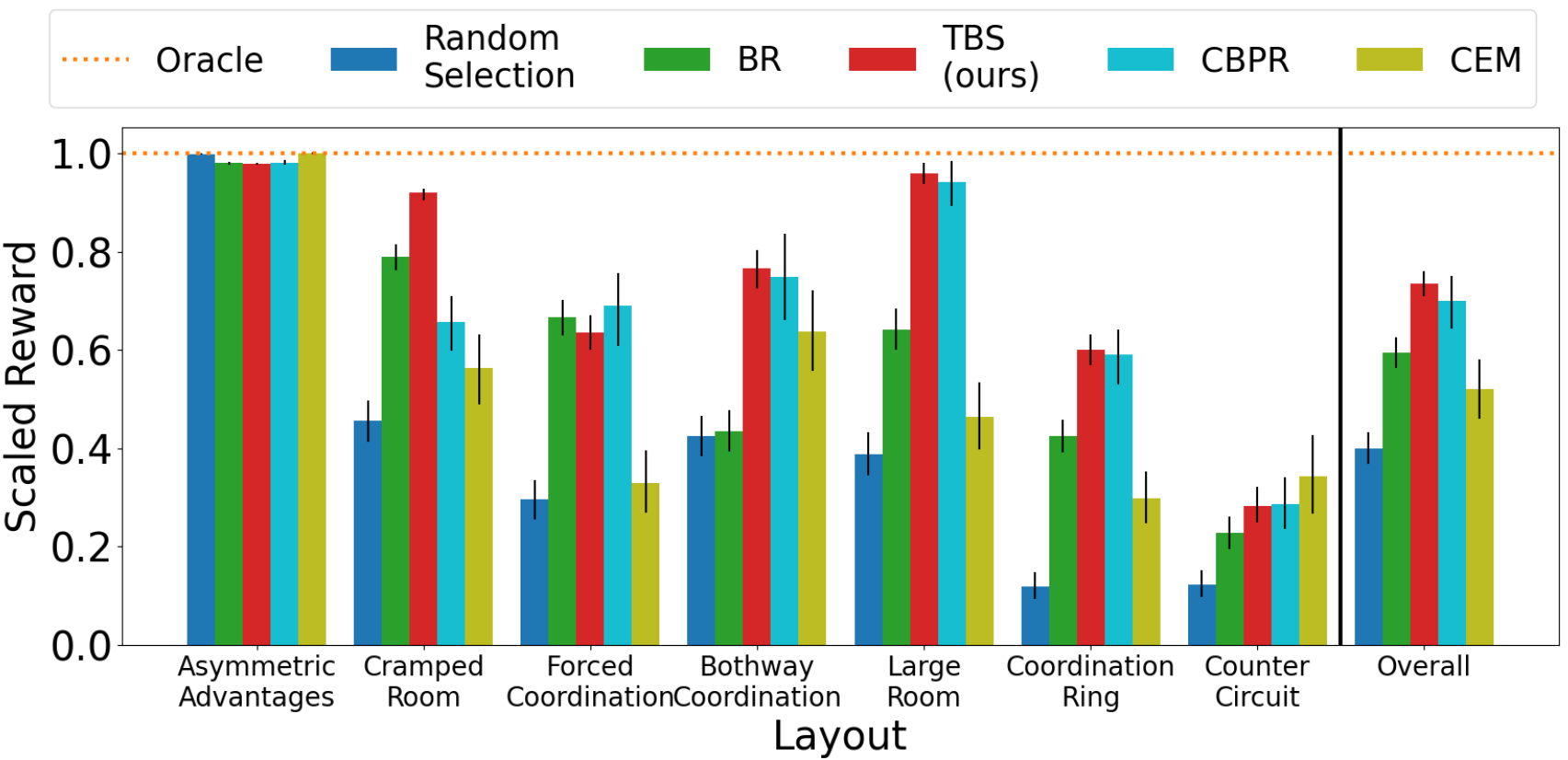}
        \caption{Fully Observable}
        \Description{Displays results for 7 fully observable Simple Soup environments. Described in text}
    \end{subfigure}%
    \hspace{7ex}
    \begin{subfigure}[b]{0.48\textwidth}\vskip 0pt
        \centering
        \includegraphics[width=1.\textwidth]{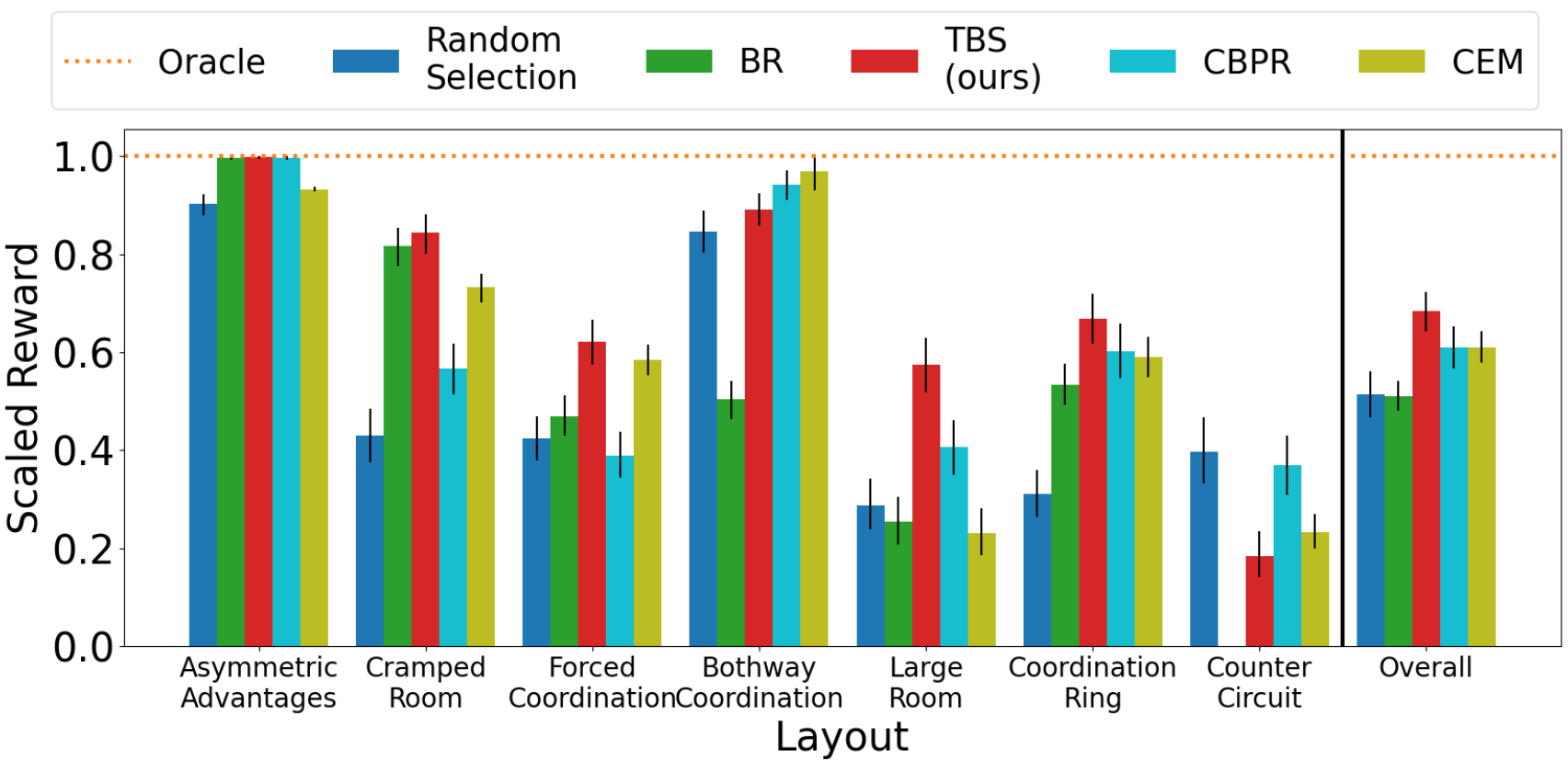}
        \caption{Partially Observable}
        \Description{Displays results for 7 partially observable Simple Soup environments. Described in text}
        
    \end{subfigure}
    \caption{ZSC Performance of different methods in 7 Onion Soup layouts in fully observable (left) and partially observable (right) settings. The final “Overall” bar in each plot denotes the average performance across all layouts.
    Error bars indicate bootstrapped 95\% confidence intervals capturing variability over different evaluation seeds. Our TBS method equals or improves on the baseline BR method in almost all layouts.} 
    \label{fig:v6_sigmoid}
\end{figure*}

\section{Experiments}\label{Experiments}

In this section, we evaluate the proposed TBS against several baselines in the Overcooked environment, focusing on ZSC.

\subsection{Experiment Setup}

\textbf{Environment.}~ We evaluate our method in the Overcooked-AI environment~\cite{overcooked-ai}, 
a widely used benchmark for ZSC. 
Specifically, we use the \textit{Onion Soup} scenario, 
in which two agents must cooperate to prepare and deliver onion soup dishes, 
mimicking the gameplay dynamics of the commercial game \textit{Overcooked}. 
Preparing a dish requires placing three onions into a pot, 
waiting 20 frames for the soup to cook, retrieving the cooked soup with a plate, 
and serving it at the delivery counter. 
Each successful delivery yields a reward of 20, and the objective is to maximize the number of deliveries within a 400-frame episode. Each agent receives a spatial observation encoding the positions and states of key items and agents, and selects actions from movement directions and a context-dependent \textit{interact} action for picking up, placing, and delivering items.

We assess performance across seven distinct layouts, each designed to elicit different coordination challenges and cooperative behaviors. For example, the \textit{Forced Coordination} layout requires the red agent to pass onions across the counter to the blue agent, whereas the \textit{Cramped Room} layout emphasizes fine-grained movement coordination in tight spaces. 
All layouts are evaluated under both fully observable and partially observable conditions. The seven layouts are illustrated in Fig.~\ref{fig:layouts} and described in detail in Appendix~\ref{app:e}.

\textbf{Concept Definition.} ~As discussed in Sec.~\ref{sec:tom}, ToM models predict concepts that capture the high-level intentions of the partner. 
In the Overcooked environment, each concept corresponds to a semantically meaningful high-level action that reflects the agent’s current subgoal 
(e.g., ``onion\_pickup\_from\_pile,'' ``onion\_drop\_in\_pot,'' ``dish\_delivery''). 
These concepts represent distinct uses of the \textit{interact} button (e.g., picking up, placing, or delivering items) 
and provide interpretable supervision signals for learning partner intent. 
A detailed description of all concept categories used in our experiments is provided in Appendix~\ref{app:c}.

\textbf{Baselines.}~ (i) \textit{Random Selection}: randomly selects an agent from our training pool to play out the entire episode, used to assess the effectiveness of adaptive selection. (ii) \textit{Best-Response} (BR): trains a single best-response agent to coordinate with the entire training pool. (iii) \textit{Oracle}: uses the exact policy that was co-trained with the unknown partner, essentially measuring $J_{SP}$ of the base algorithm,  and thus serves as an upper bound. (iv) \textit{CEM}: Uses the cross-entropy metric described in~\cite{tsf} to choose an agent from the training pool to cooperate with the partner. (v) \textit{CBPR}: Uses the intra-episode bayesian belief update proposed in~\cite{cbpr} to choose a specialist best-response agent to cooperate with the teammate.

Both the baselines and the proposed TBS require constructing an agent pool, which we create by training 10 agents. \textit{Random Selection} and \textit{CEM} do not require additional RL training, whereas \textit{Best-Response}, \textit{CBPR} and \textit{TBS} involve further training. Specifically, \textit{Best-Response} trains a single cooperator agent by pairing it with one of the 10 agents in the pool, randomly selected at the beginning of each training episode, and \textit{CBPR} and \textit{TBS} pair each cooperator agent with agents randomly selected from its corresponding strategy cluster. We provide the training details in Appendix \ref{app:d}.

\textbf{Evaluation.}~ We evaluate the any-play performance, $J_{\text{inter-XP}}$, of the proposed TBS and the baselines. 
Specifically, this measures performance when playing with held-out agents that are independently trained using VDN with randomly shaped rewards. Note that the held-out agents follow different strategies, as they are trained with distinct reward functions. For this purpose, we generate 10 different held-out agents and use them for evaluation. 

Code and experiment scripts are available at: \url{https://github.com/andrewni2002/ToMZSC}


\subsection{Experimental Results}\label{sec:results}

We provide results under both fully and partially observable settings, 
as shown in Fig.~\ref{fig:v6_sigmoid}(a) and (b), respectively. The figures also report the average performance across the considered layouts. Across both settings, TBS achieves higher average performance than the baseline methods, demonstrating superior adaptability  in ZSC.

In the fully observable setting, the proposed TBS method matches or exceeds the performance of the BR baseline in six out of seven layouts. 
While the BR agent tends to learn a static generalist policy that fails to adapt to unseen teammates, 
TBS leverages explicit ToM-based adaptation through its ensemble of specialized BR policies, 
enabling rapid identification of the teammate’s strategy and effective adaptation to it, 
which leads to higher ZSC performance overall. In the partially observable setting, the overall ZSC scores decrease due to the increased difficulty of coordination under limited observability. 
Nevertheless, TBS matches or outperforms the BR baseline across all layouts, 
demonstrating its robustness and adaptability even in partially observable environments.

\begin{figure*}[t]
    \centering
    \begin{subfigure}[b]{0.2\textwidth}\vskip 0pt
        \centering
        \includegraphics[width=\textwidth]{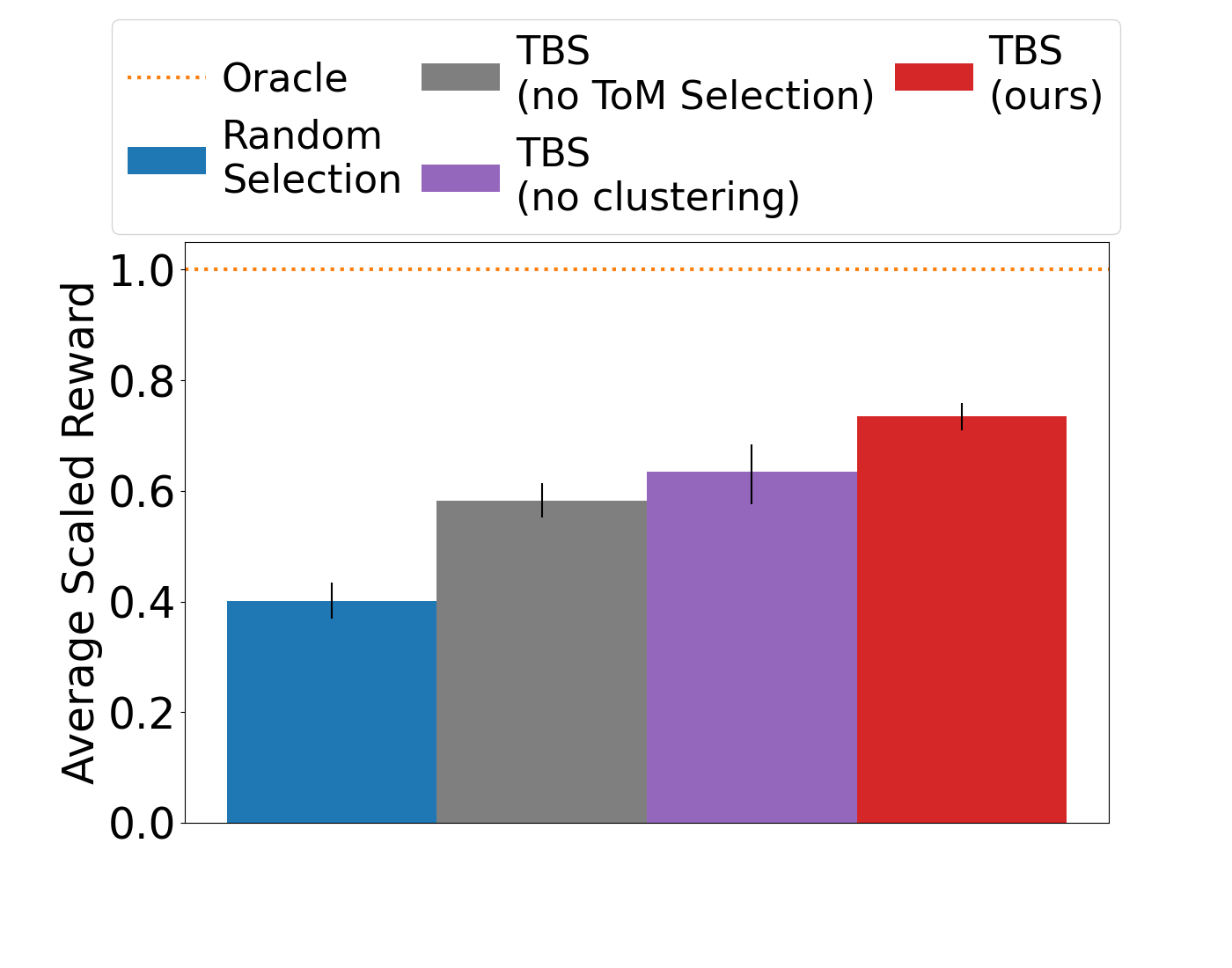}
        \caption{Ablation main components}
        \label{fig:contribution}
        \Description{Fully described in text}
    \end{subfigure}
\begin{subfigure}[b]{0.225\textwidth}\vskip 0pt
        \centering
        \includegraphics[width=\linewidth]{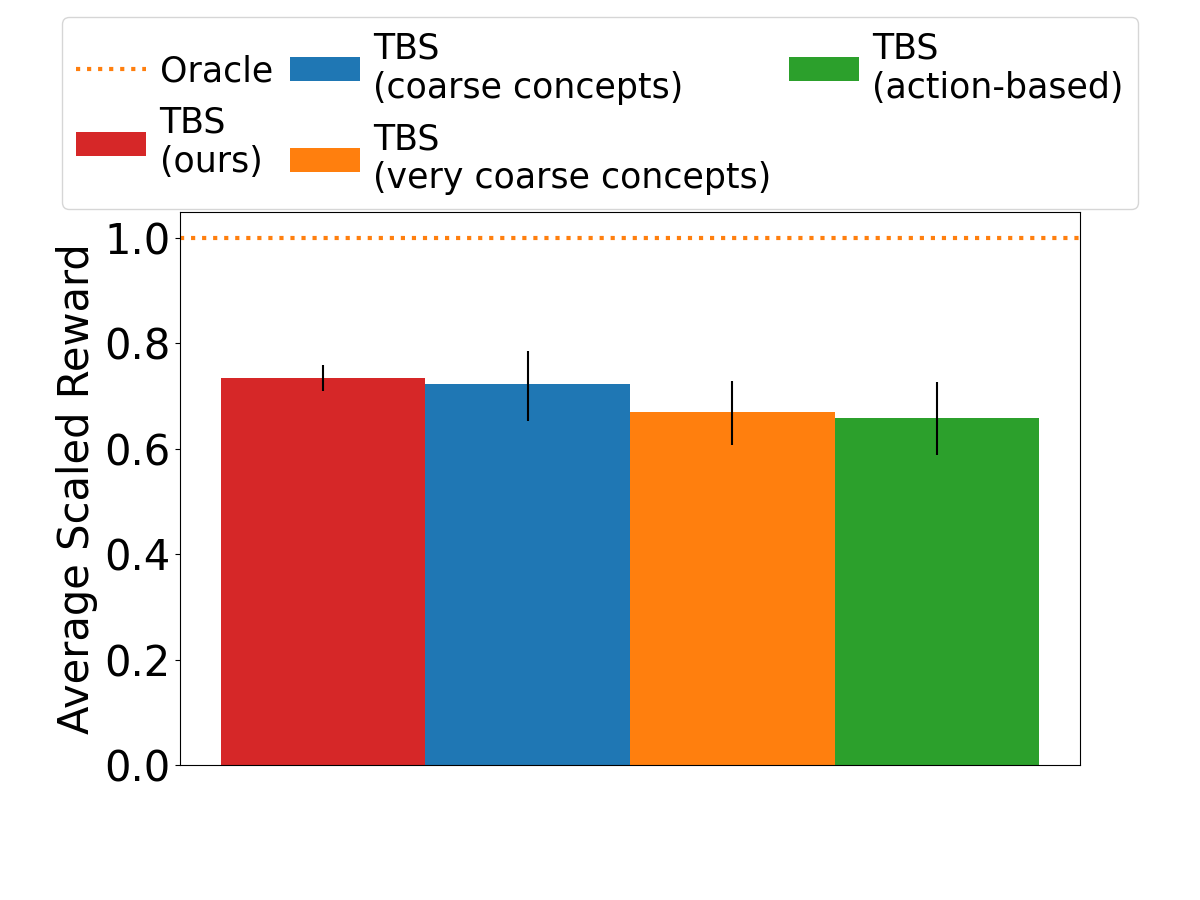}
        \caption{Ablation concept set}
        \label{fig:concept_ablation}
        \Description{Fully described in text}
    \end{subfigure}
    \begin{subfigure}[b]{0.272\textwidth}\vskip 0pt
        \centering
        \includegraphics[width=\textwidth]{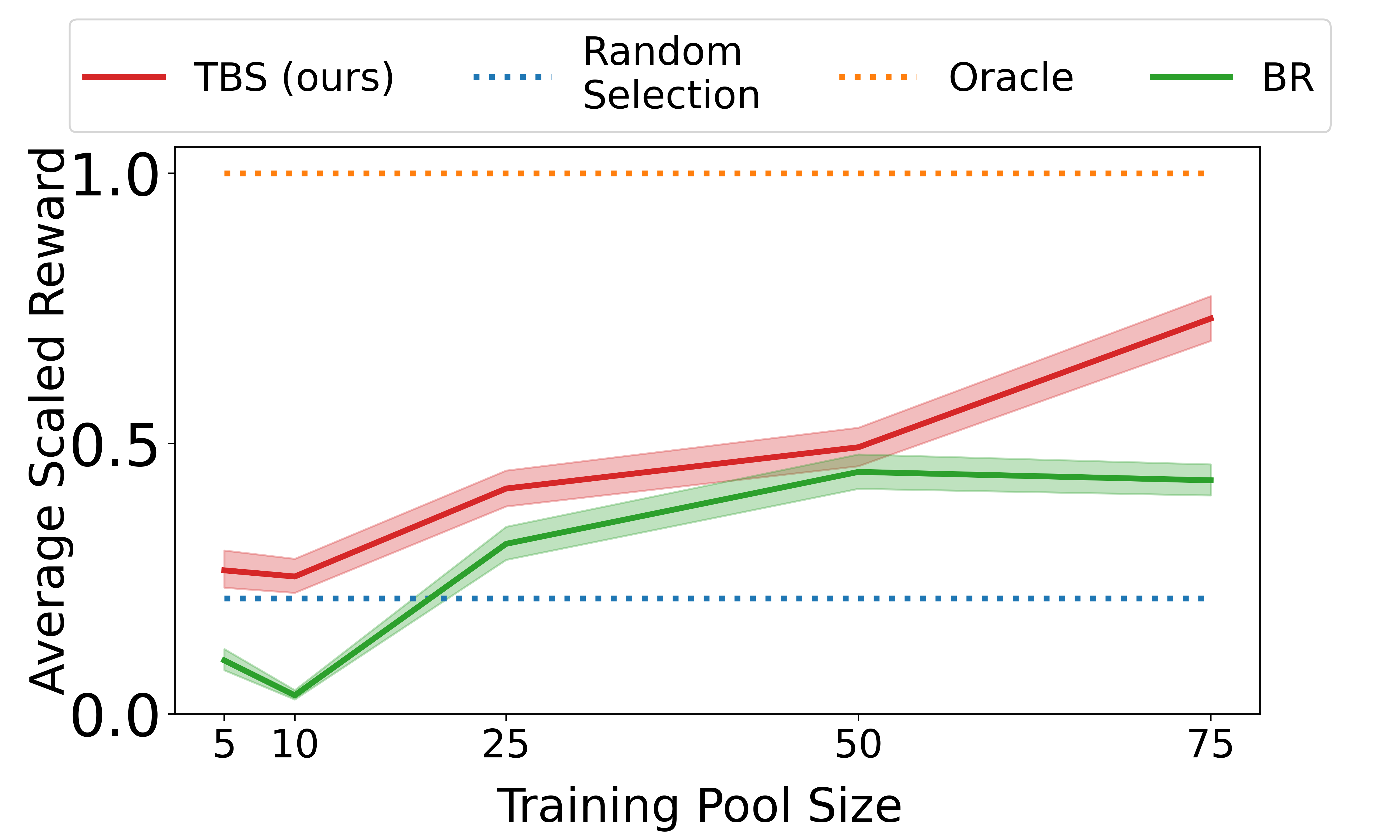}
        \caption{Ablation training pool size}
        \label{fig:nagents_ablation}
        \Description{Fully described in text}
    \end{subfigure}%
    \begin{subfigure}[b]{0.272\textwidth}\vskip 0pt
        \centering
        \includegraphics[width=\textwidth]{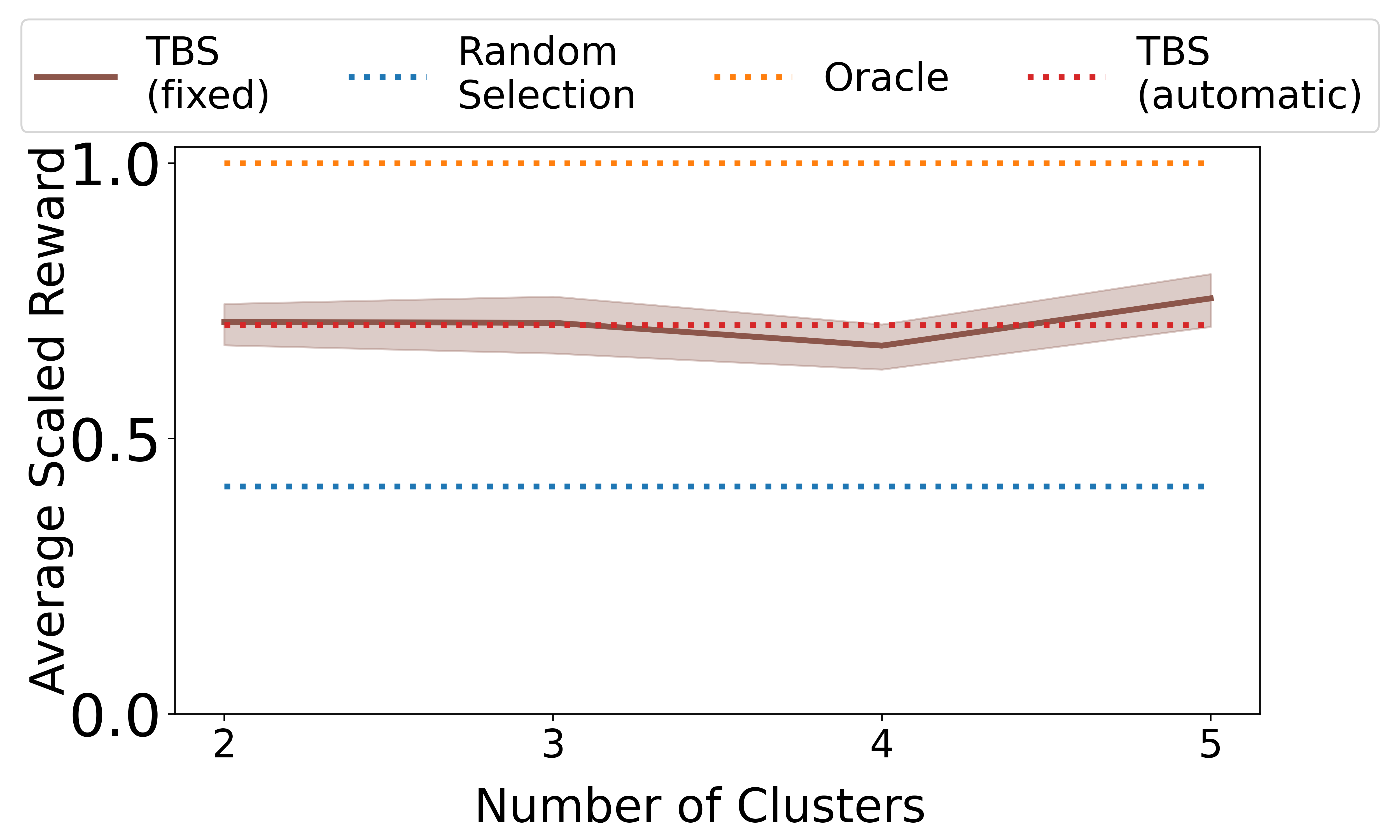}
        \caption{Ablation number of clusters}
        \Description{Fully described in text}
        \label{fig:ncluster_ablation}
    
    \end{subfigure}
    
    \caption{Experiments ablating the main components of TBS (left), the number of agents in the training pool (middle) and the number of strategy clusters when using spectral clustering with a fixed number of clusters (right). Error bars and shaded regions indicate bootstrapped 95\% confidence intervals capturing variability over different evaluation seeds \textbf{Left:} Average ZSC performance of different methods over all 7 onion soup layouts. Both clustering and ToM-based strategy selection improve ZSC performance.  \textbf{Middle:} ZSC performance on the difficult \textit{Counter Circuit} layout as a function of the number of agents in the training pool. Larger training pools result in better ZSC performance for both TBS and BR, but TBS improves more. \textbf{Right:} Average performance of TBS with different numbers of clusters over all 7 onion soup layouts. Average scaled rewards are all around 0.7 to 0.8, indicating that our method is robust to the number of clusters.} 
    \label{fig:analysis}
\end{figure*}

\begin{figure}[t]
    \centering
    \includegraphics[width=0.7\linewidth]{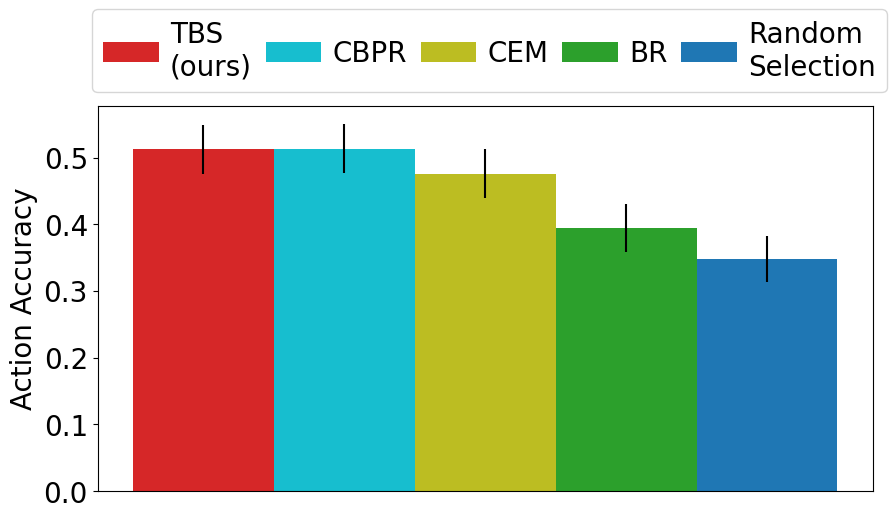}
    \caption{Frequency of different methods selecting the same action as that of the agent co-trained with the unknown teammate, averaged over 7 simple soup layouts}
    \label{fig:action_frequency}
    \Description{Displays average results over 7 fully observable Simple Soup environments. TBS and CBPR perform the best, followed by CEM, then BR, then Random Selection}
\end{figure}

In both settings, the Random Selection baseline performs significantly worse, 
highlighting the necessity of adaptive selection and the effectiveness of the ToM-based adaptive module in TBS.

\begin{figure*}[t]
  \includegraphics[width=0.7\textwidth]{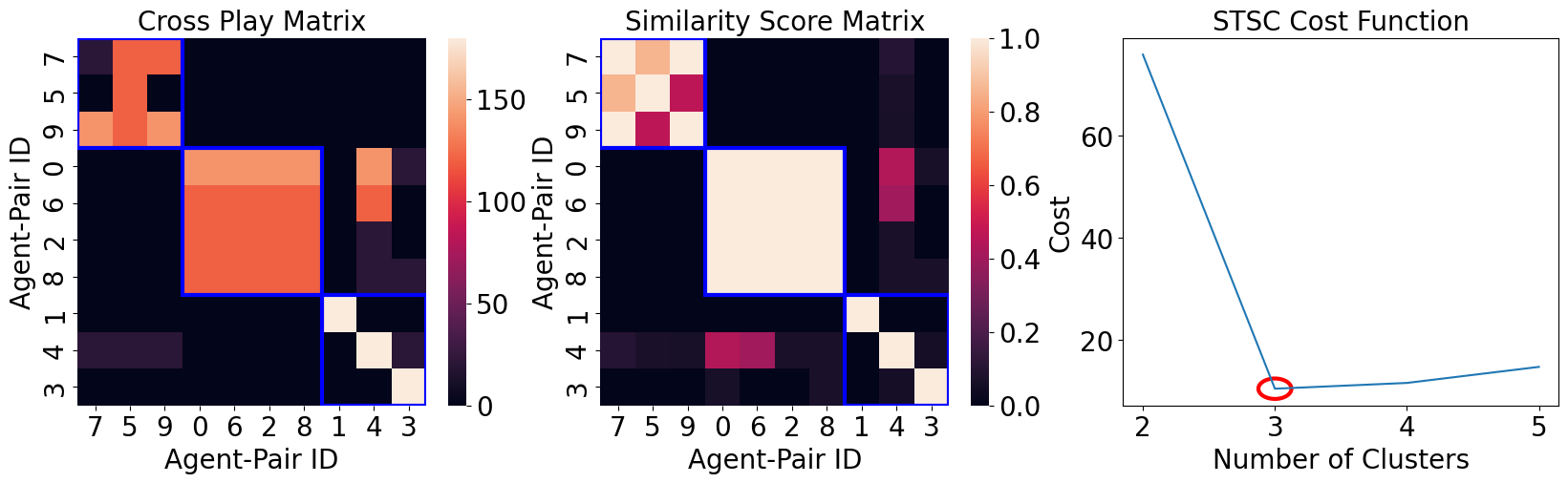}
  \caption{Self-tuning spectral clustering in the \textit{Large Room}. Left: XP matrix $J(\pi_A^1,\pi_B^2)$. Middle: similarity matrix $s(A,B)$. Right: STSC cost minimized at $k=3$ (red circle).}
  \label{fig:stsc}
  \Description{An example of self-tuning spectral clustering. The cross-play and similarity score matrices feature three clusters of 3, 4, and 3 agents. The STSC cost function decreases drastically from 2 to 3 clusters and slowly increases afterwards.}
\end{figure*}

\subsection{Analysis}

In this section, we provide several in-depth analyses for better understanding of the proposed method. Our analysis consists of two main parts. 
The first focuses on the adaptive selection module, analyzing how effectively it identifies and responds to the partner’s strategy. 
The second focuses on the partner pool and the corresponding BR policies, examining how the proposed spectral clustering operates 
and investigating the effects of the number of clusters and the size of the partner pool. 
We also include a sensitivity analysis on the concept sets used for ToM training to assess their impact on performance.

\subsubsection{Adaptive Selection.}~We first evaluate how often each method selects the same action as the oracle, defined as the co-trained agent. Since the oracle was trained jointly with the unknown partner, its policy can be viewed as a reference for the correct cooperative action at each timestep. The results are shown in Fig.~\ref{fig:action_frequency}. 
TBS achieves higher action accuracy than all baselines except CBPR.

We further perform an ablation on the adaptive selection module. 
Fig.~\ref{fig:analysis}(a) shows the results of TBS without the adaptation module and a random cross-play variant that removes both cross-play clustering and ToM-based selection. Both perform worse than TBS, indicating that the proposed ToM-based adaptive selection module is essential for effective ZSC. Additional ablation studies on window size and steps per selection are provided in Appendix~\ref{app:a}.

\subsubsection{Clustering Operation.}~We first describe how our proposed self-tuning spectral clustering operates. Fig.~\ref{fig:stsc} illustrates a walkthrough of our automatic strategy clustering procedure in the \textit{Large Room} layout. As discussed in Sec.~\ref{sec:tbs}, we first construct the cross-play matrix over the training pool and then compute the similarity score matrix $s(A,B)$ as defined in Eq.~\ref{eq:cross-play-similarity}. Finally, we apply self-tuning spectral clustering~\cite{stsc} on the similarity matrix
, minimizing the cost function from Eq.~\ref{eq:stsc} to determine the number of clusters. 

As shown in the left and middle panels of Fig.~\ref{fig:stsc}, three clusters (boxed in blue) are identified, indicating that the policies within each blue box exhibit similar behaviors. The number of clusters is determined to be three, as shown in the rightmost panel of Fig.~\ref{fig:stsc}, where the STSC cost function reaches its minimum when $k = 3$. It is observed that agents with relatively higher cross-play performance tend to be grouped together, 
which aligns with the intended objective of clustering behaviorally similar strategies. 
Additional clustering results for other layouts are provided in Appendix~\ref{app:b}.


In addition, we conduct an ablation on the clustering component. Fig.~\ref{fig:analysis}(a) shows the result of TBS without clustering. Removing the clustering step leads to a clear drop in performance, highlighting
its importance in the overall framework.

\subsubsection{Size of Partner Pool}

Having a diverse set of partners is one of the most important factors for achieving effective ZSC---exposing agents to a wide range of partners during training improves their ability to collaborate with unseen partners. Therefore, it is crucial to leverage a sufficiently large partner population within a given computational budget, such as memory and compute capacity. 
As observed in the toy example in Sec.~\ref{sec:motivation}, performance generally improves as the partner pool size increases. 
However, while the BR baseline struggles to effectively leverage larger populations, TBS can continue to benefit from increased partner diversity. To verify this observation in a more realistic setting, we conduct experiments in the Overcooked environment, varying the size of the partner pool between 5 and 75 in the most challenging layout, \textit{Counter Circuit}, 
and report the average ZSC performance of both TBS and the BR baseline on 25 held-out evaluation agents.

The corresponding results are shown in Fig.~\ref{fig:analysis}(c). 
It is observed that (i) TBS consistently outperforms BR in ZSC performance, in line with our previous findings; 
(ii) the ZSC performance of both methods increases with the training population size, which aligns with intuition and prior studies~\cite{gamma,pecan,mep,fcp}; 
and (iii) TBS benefits more from larger partner populations, as it can effectively exploit agent diversity to construct an adaptive ensemble, 
whereas BR learns a static, generalist policy that limits its performance gains from increased population size. 
This aligns with the result observed in the simple toy example, described in Sec~\ref{sec:motivation}.

\subsubsection{Number of Clusters}

While we have used self-tuning spectral clustering~\cite{stsc}, which automatically determines the number of clusters, we can also treat the number of strategy clusters as a tunable hyperparameter. In this analysis, instead of using self-tuning spectral clustering, we manually set the number of clusters $k$ and then perform vanilla spectral clustering to construct the strategy clusters. We train a population of 10 training agents and 10 held-out evaluation agents using VDN, 
vary $k$ between 2 and 5, and report the average ZSC performance of our method over the seven fully observable \textit{Simple Soup} layouts described above. 

The results are shown in Fig.~\ref{fig:analysis}(d). 
TBS (fixed) indicates TBS with a preset number of clusters, whereas TBS (automatic) refers to the original method using self-tuning spectral clustering. 
The average performance of TBS (fixed) stays around 0.7–0.8, indicating that our method is robust to the choice of $k$. In addition, using self-tuning spectral clustering eliminates the need for manual hyperparameter tuning while maintaining comparable performance.

\subsubsection{Concept Set}

The ToM models in TBS predicts a partner’s intention as a distribution over a predefined set of concepts. To understand how the choice of this concept set affects performance, we investigate the sensitivity of TBS to different levels of concept granularity.
While we have used a granular concept set containing a total of 44 concepts for TBS, in this section we explore using TBS with less granular and different kinds of concept sets. Specifically, the TBS (coarse concepts) method uses a coarser concept set of size 11 that does not distingush between key items at different locations. The TBS (very coarse concepts) uses an even lower resolution concept set containing only 6 concepts. Finally the TBS (action-based) method directly uses the 6 low-level actions as its concept set.

The average ZSC performance of each method over 7 fully observable Onion Soup layouts is displayed in Fig.~\ref{fig:concept_ablation}. The average ZSC performances of the four methods are similar, indicating the robustness of TBS to concept set specification. However, we also see that high-level intention-based concept sets outperform low-level action based concept sets, and that more granular concept sets outperform less granular ones, as concept sets that are less granular and less aligned with the problem of high-level coordination are more likely to conflate distinct strategies and choose the wrong specialist best-response agent.

\section{Conclusion}

In this work, we presented TBS, a Theory-of-Mind-based Best Response Selection framework designed to enhance zero-shot coordination with novel partners. Unlike conventional population-based approaches that rely on a single static best-response policy, TBS leverages behavioral clustering and ToM-guided policy selection to achieve adaptive collaboration. During evaluation, the agent infers its partner’s behavioral intent and selects the most compatible best-response policy in real time, enabling robust adaptation to unseen strategies. Empirical evaluations in the Overcooked environment demonstrate that TBS significantly improves coordination performance across diverse conditions, underscoring the potential of explicit Theory-of-Mind reasoning as a general mechanism for zero-shot coordination.

\textbf{Limitation}~One limitation of our work is the lack of theoretical analysis on coordination and generalization. We leave this for future work. In addition, our method depends on the specific concept sets used, and underspecification of concept sets could lead to different strategies being conflated by our ToM-based ensemble coordinator.




\begin{acks}
This work was supported in part by the DARPA EMHAT Program under Agreement No. HR00112490409, the ONR Award No. N00014-23-1-2840, the Okawa Foundation Research Grant, and the NSF CAREER Award No. 2145077.
\end{acks}



\clearpage
\bibliographystyle{ACM-Reference-Format} 
\bibliography{refs}

\clearpage

\appendix

\section{Additional Ablation Studies}\label{app:a}

    \begin{figure}
        \centering
        \includegraphics[width=\linewidth]{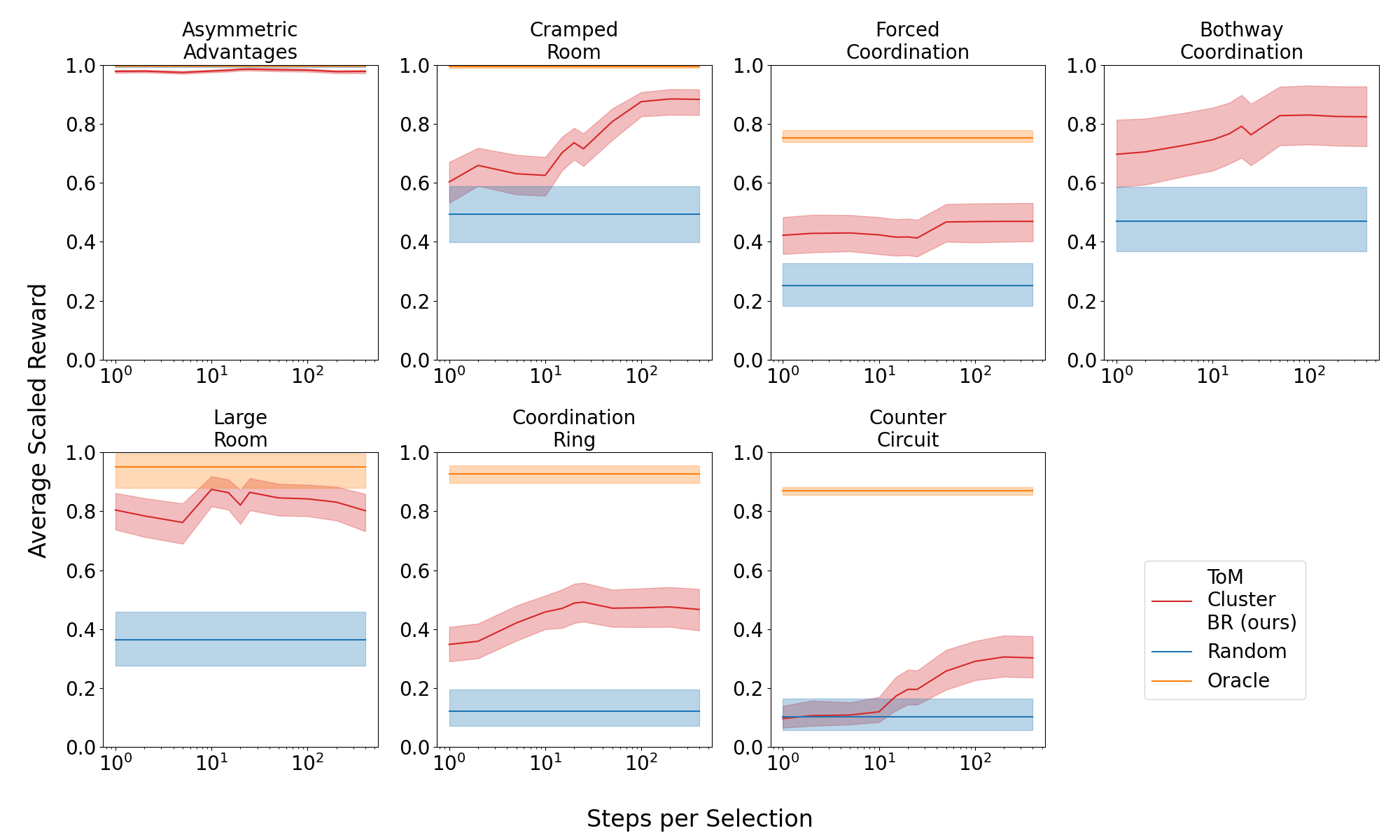}
        \caption{Ablation results varying the number of timesteps in the rolling window for ToM distance computation. Note that the x-axis is in log-scale for better visualization. Shaded regions represent bootstrapped 95\% confidence intervals. Performance generally increases for larger window sizes, as TBS can more accurately determine the teammate's strategy.}
        \label{fig:window-size-ablation}
    \end{figure}

\subsection{Window Size}

During online adaptation to new partners (see Sec.~\ref{sec:onlineadaptation}), TBS selects the most appropriate cluster BR agent based on ToM outputs accumulated over all previous timesteps. This section examines the performance of TBS when only recent timesteps are used for ToM-based selection. Specifically, we modify the ToM distance computation to consider a rolling window of the most recent $w$ timesteps, and vary $w$ between 1 and 400. That is, we modify the metric for BR agent selection in Eq.~\ref{eq:onlineadapt} as follows:
\begin{equation}
        D_i=\sum_{t'=\max(0,t-w)}^t\sum_{j=1}^MD_{\text{KL}}(\text{Bernoulli}(\hat{c}_{t',i}^1[j])||\text{Bernoulli}(c_{t'}^1[j]))
\end{equation}

Intuitively, smaller window sizes will make it more difficult to accurately identify the most appropriate cluster BR agent for cooperating with the teammate, but provide greater flexibility when the teammate’s behavior shifts between strategies, which may occur, for example, when cooperating with adaptive agents.


We vary the window size $w$ between 1 and 400, and report the average performance over the seven fully observable Onion Soup layouts described above. The results are shown in Fig.~\ref{fig:window-size-ablation}. Overall, increasing the window size improves performance in the Onion Soup environments, as TBS is able to more accurately identify the most suitable BR agent.

\subsection{Steps per Selection}

During online adaptation to new partners (see Sec.~\ref{sec:onlineadaptation}), TBS selects the most appropriate cluster BR agent at every timestep. In this section, we examine the performance of TBS when selection events are less frequent. We modify TBS to select a new BR agent $A_t$ every $n$ timesteps, and continue executing that agent's policy for the next $n$ timesteps before selecting a new agent. Intuitively, decreasing the selection frequency reduces the adaptiveness of our method and thus lowers ZSC performance.

\begin{figure}
        \centering
        \includegraphics[width=\linewidth]{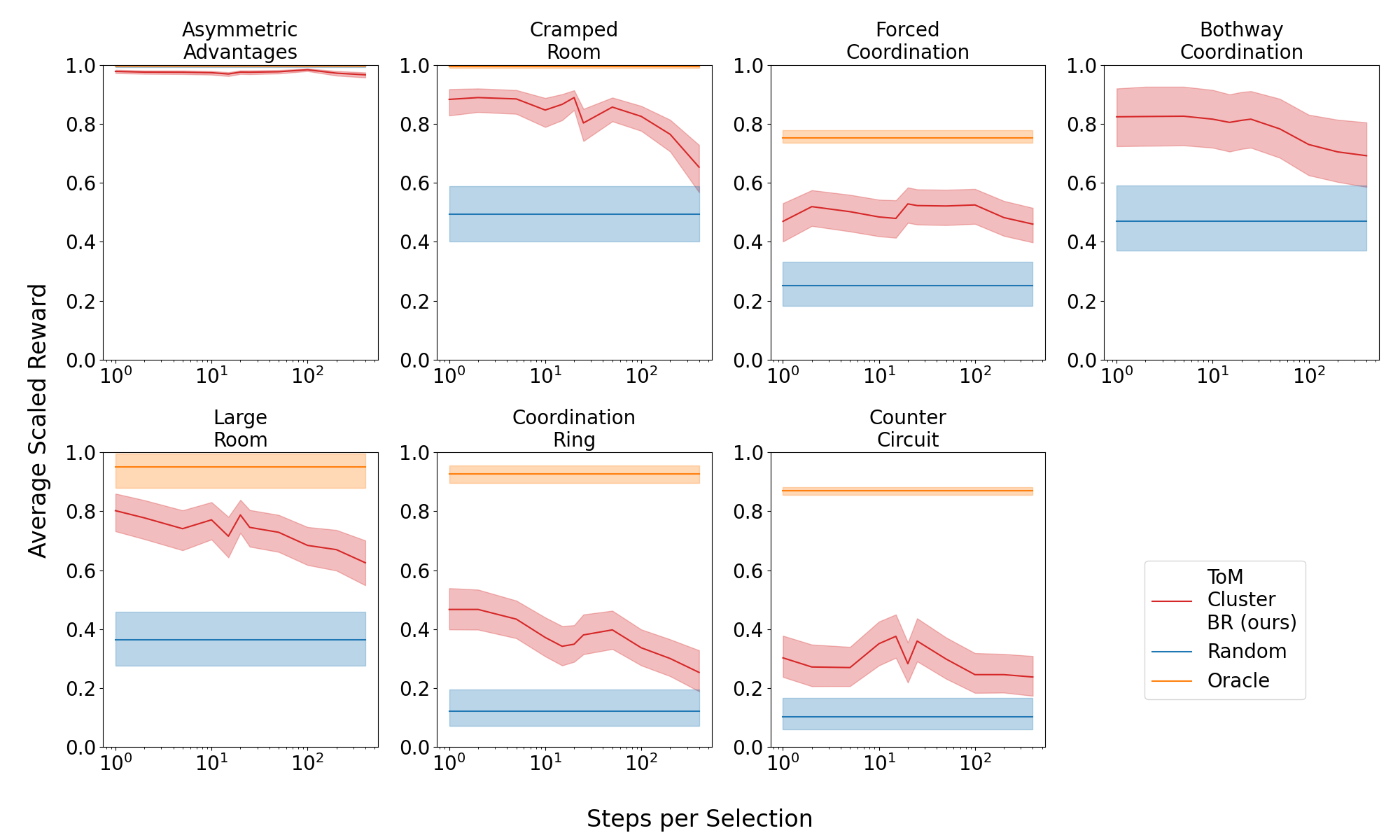}
        \caption{Ablation results varying the ToM-based agent selection frequency. Note that the x-axis is in log-scale for better visualization. Shaded regions represent bootstrapped 95\% confidence intervals. Performance generally decreases as the selection frequency decreases (i.e. steps per selection event increases) due to decreased adaptiveness.}
        \label{fig:steps-ablation}
    \end{figure}

We vary the number of steps per selection event between 1 and 400, and report the average performance over the seven fully observable Onion Soup layouts described above. The results are shown in Fig.~\ref{fig:steps-ablation}. Overall, increasing the number of steps per selection event size decreases ZSC performance in the Onion Soup environments, as it reduces the ability of TBS to adapt to new partners.



\section{Clustering}\label{app:b}

In this section we provide supporting figures related to our automatic cross-play based strategy clustering method. We use self-tuning spectral clustering on the cross-play similarity score matrix defined in Eq.~\ref{eq:cross-play-similarity} to automatically determine the number of clusters and group agents into clusters.

\subsection{Cross Play Matrices}

\begin{figure}
        \centering
        \includegraphics[width=\linewidth]{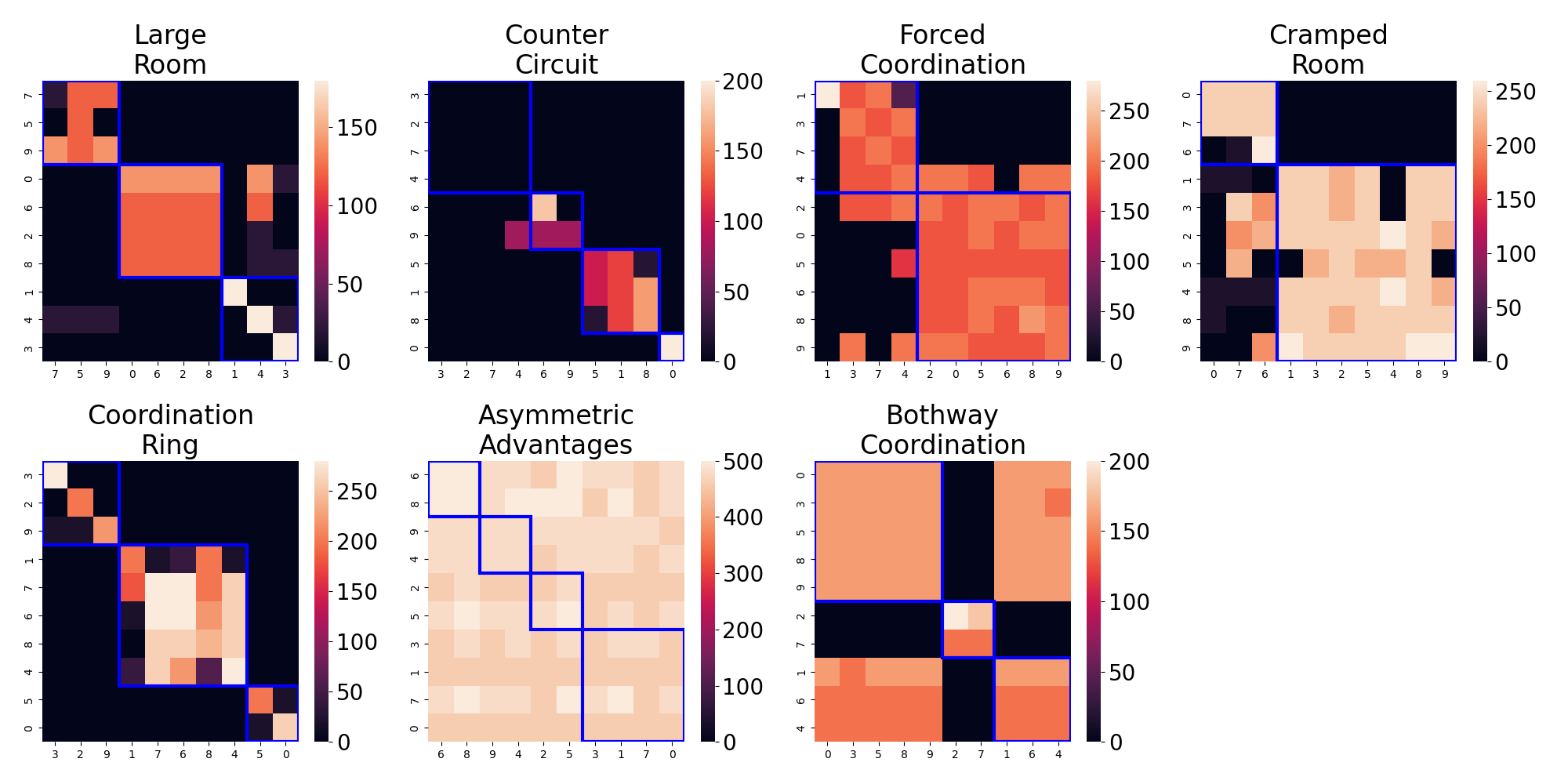}
        \caption{Cross-play matrices for training agent pools in the Simple Soup environment layouts. Clusters identified by self-tuning spectral clustering are outlined in blue.}
        \label{fig:crossplay}
    \end{figure}

\begin{figure}
        \centering
        \includegraphics[width=\linewidth]{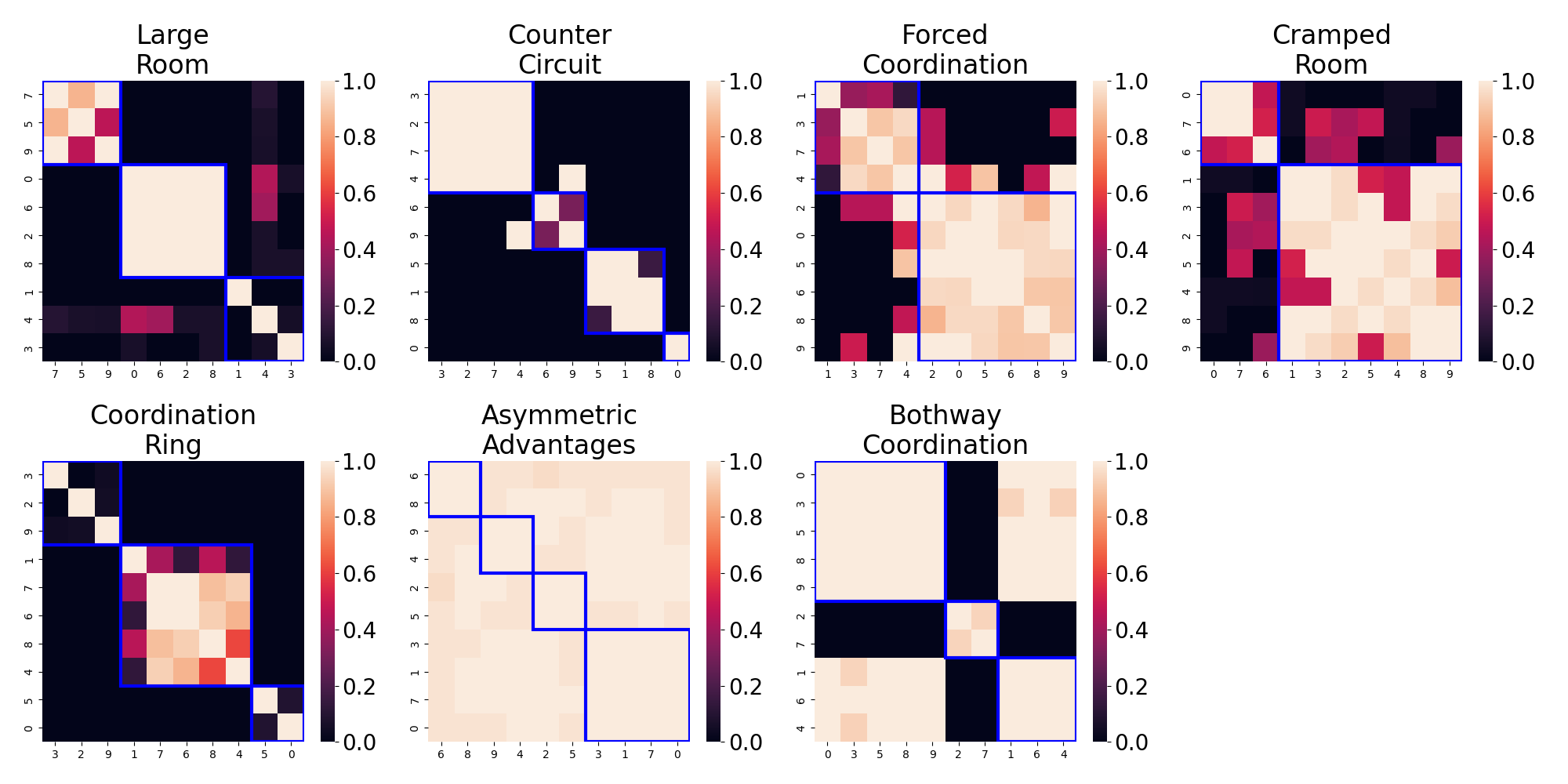}
        \caption{Similarity score matrices for training agent pools in the Simple Soup environment layouts. Clusters identified by self-tuning spectral clustering are outlined in blue.}
        \label{fig:similarity}
    \end{figure}

In this section we provide heatmaps of the cross play matrices $(X)_{ij}=J(\pi_i^1,\pi_j^2)$ and cross play similarity score matrices $(S)_{ij}=s(i,j)=\frac{J(\pi_i^1,\pi_j^2)+J(\pi_j^1,\pi_i^2)}{J(\pi_i^1,\pi_i^2)+J(\pi_j^1,\pi_j^2)}$ along with the clusters determined by self-tuning spectral clustering~\cite{stsc}. Note that in practice, we clamp $s(i,j)$ to the range $[0,1]$ as sometimes the cross-play reward might be higher than the self-play reward. In addition, when both cross-play and self-play rewards are 0, making the similarity score undefined, then we say that the similarity score is 1, since the agent pairs are similar in that they've both failed to solve the environment. Finally, we add $1\times10^{-4}$ to all similarity scores to avoid running into singular matrices during spectral clustering. The resulting cross-play and similarity matrices are visualized in Figs.~\ref{fig:crossplay} and \ref{fig:similarity}, respectively.

\section{Concept Theory of Mind}\label{app:c}

\subsection{Concept Sets}
In this section we detail the exact high-level concepts used in our experiments. Our concepts are intentions over high-level actions, where these high-level actions are generally of the form ``pickup/drop [item] off/on [tile]'' and correspond to the different possible uses of the ``interact'' button in Overcooked. 

For the Simple Soup environments, the possible actions include picking up and placing down onions, plates, or soup dishes, and the possible tiles include onion/plate piles, empty counters, pots, and serve stations. Therefore, our high-level action set is 
\begin{enumerate}
    \item onion\_pickup\_from\_pile, 
    \item plate\_pickup\_from\_pile,
    \item dish\_pickup\_from\_pot,
    \item onion\_pickup\_from\_counter, 
    \item plate\_pickup\_from\_counter,
    \item dish\_pickup\_from\_counter,
    \item onion\_drop\_in\_pot,
    \item onion\_drop\_on\_counter,
    \item plate\_drop\_on\_counter,
    \item dish\_drop\_on\_counter,
    \item dish\_delivery
\end{enumerate} 

In addition, we distinguish between up to 4 special tiles of each type. Therefore, putting an onion in one pot is a different concept from putting an onion in a different pot. For placing objects on counters, we define empty counters in the middle of the layout (relevant for \textit{Counter Circuit}, \textit{Bothway Coordination}, \textit{Coord Ring}, and \textit{Forced Coord}) as special, and all other counters as identical.


\section{Training Details}\label{app:d}

\subsection{RL Hyperparameters}

In this section we detail the hyperparameters used for RL training. As mentioned previously, we use VDN with random reward shaping for the base RL algorithm, and DQN for training best-response agents. We use lambda-targets for our Q targets in both VDN and DQN. Our RL hyperparameters overall follow the JaxMARL implementation \cite{jaxmarl} and are as follows:

\begin{table}
  \caption{Simple Soup Hyperparameters}
  \label{simple-soup-rl}
  \centering
  \begin{tabular}{lll}
    Hyperparameter     & VDN (Training Pool)   & DQN (BR)\\
    \midrule
    Timesteps & $5\times10^6$  & $4\times10^7$     \\
    Num Parallel Envs & 64 & 64      \\
    Num Steps     & 16 & 100  \\
    Epsilon Anneal     & 0.2 & 0.2  \\
    Num Minibatches     & 16 & 16  \\
    Num Epochs     & 4 & 4  \\
    LR     & 0.000075 & 0.000075  \\
    LR Decay     & Linear & Linear \\
    Lambda & 0.5 & 0.5 \\ 
    Gamma & 0.99 & 0.99 \\ 
    Reward Shaping Horizon & $4\times10^6$ & $3\times10^7$\\
    \bottomrule
  \end{tabular}
\end{table}


We ran RL on Quadro RTX 6000 GPUs, and overall, we found that for each layout it took around 2 GPU-hours to train the training agent population, and around 1-2 hours to train each best-response agent.

\subsection{Random Reward Shaping}

In addition to the annealed reward shaping for improving the sample-efficiency of RL, we also add constant, randomly sampled reward shaping to guide the agents towards diverse behaviors. For each environment we define a set of game events similar to those used by the default annealed reward shaping. Then for each agent in each RL run, we sample a random normally distributed reward coefficient for each game event. During the run, every time this agent triggers a game event, it receives the corresponding random reward in addition to the sparse reward and the annealed shaped reward. To encourage more diverse behaviors, we scale the magnitude of the random rewards to be approximately inversely proportional to frequency with which we expect each game event to be triggered. For example, since agents in the simple soup environment must put three onions in the pot for each soup dish and pick up the cooked soup once, the random reward assigned to picking up plates will have magnitude 3x larger than that of placing onions in the pot.

The specific game events and random reward magnitudes are listed below 

\begin{itemize}
    \item Simple Soup
    \begin{enumerate}
        \item Place onion in pot: 0.15, 
        \item Pickup plate: 0.5
        \item Pickup soup: 0.5
        \item Pickup item from counter: 0.15
        \item Drop item on counter: 0.15
        \item Deliver soup: 0.5
    \end{enumerate}
\end{itemize}

\subsection{ToM Model Training}

This section details our training process for the ToM models, and specifically, how ground-truth intention concepts are constructed from sampled trajectories.
When training our ToM models, we add to each state-action-reward tuple $(s_t,a_t,r_t)$ in our MDP an interaction vector $I_t$ which encodes the different ways an agent can interact with the environment beyond simple movements (i.e. actions accessible through the ``interact'' button. See previous section for categorization of possible high-level actions), where $I_t^1[i]=1$ when agent 1 is currently performing interact action $i$ (e.g. pickup onion from onion pile) and 0 otherwise. $I_t^2$ is defined similarly for agent 2. Then for each timestep $t$, we define the next interact timestep of agent 1 $t_{\text{next}}^1=\min\{t'|t\le t'\le T \text{ and } \sum_iI_{t}^1[i]>0\}$ as the next timestep in which agent 1 performs some interact action. Note that $t^1_{\text{next}}$ may not exist when $t$ is close to $T$, as the agent does not have time to take an ``interact'' action before the end of the episode. $t^2_{\text{next}}$ is defined similarly. We then define the ground-truth intention concept $c_t^1=I_{t_{\text{next}}^1}$, intuitively the next high-level action taken, when $t_{\text{next}}^1$ exists and $c_t^1[i]=0$ for all $i$ otherwise. $c_t^2$ is defined similarly. Note that this process assumes that agents do not change their minds in the middle of performing a high-level action, and that agents have sufficient grasp over game controls so their actions actually reflect their intentions, instead of intending to do one action but accidentally performing another action.

After constructing the ground-truth concepts, we train our ToM model to predict the intentions of agent 1 $c_t^1$ given the observation history of agent 2: $[o_t^2,o_{t-1}^2,\cdots,o_1^2]$. Specifically, since our concepts are binary-valued, we minimize the binary cross entropy loss 
    
    \begin{equation}
    \begin{split}
        \mathcal{L}(\theta_{\text{ToM},i})=&\mathbb{E}_{A\sim U[\mathcal{C}_i]}\mathbb{E}_{\tau\sim[\pi_A^1,\pi_{C_i}^2]}\\&\left[\frac{1}{T}\sum_{t=1}^T\text{BCE}(\text{ToM}(o_t^2,\cdots,o_1^2;\theta_{\text{ToM},i}),c_t^1)\right]
    \end{split}
    \end{equation}

     where $\text{BCE}(\hat c_t^1,c_t^1)=\sum_ic_t^1[i]\log\hat c_t^1[i]+(1-c_t^1[i])\log(1-\hat c_t^1[i])$

The 3rd observer is trained similarly, but using all trajectories across all clusters.
    \begin{equation}
    \begin{split}
        \mathcal{L}(\theta_{\text{3rdObs}})=&\mathbb{E}_{i,j\overset{\mathrm{iid}}{\sim} U[k]}\mathbb{E}_{A\sim U[\mathcal{C}_i]}\mathbb{E}_{\tau\sim[\pi_A^1,\pi_{C_j}^2]}\\&\left[\frac{1}{T}\sum_{t=1}^T\text{BCE}(\text{ToM}(o_t^2,\cdots,o_1^2;\theta_{\text{3rdObs}}),c_t^1)\right]
        \end{split}
    \end{equation}

We trained our ToM models on Quadro RTX 6000 GPUs, and overall found that training the ToM models for a single layout takes around 1-3 GPU-hours. In addition, ZSC evaluation was done on cpus, taking about 1 hour per layout. 




\section{Layouts}\label{app:e}

In this section we describe the environment layouts used in this work.

\subsection{Onion Soup}

We use the \textit{Counter Circuit}, \textit{Asymmetric Advantages}, \textit{Forced Coordination},  \textit{Cramped Room} layouts proposed in \cite{overcooked-ai}. In addition, we take the \textit{Bothway Coordination} layout proposed in \cite{zsceval}. While we initially also tried to use the \textit{Blocked Corridor} and \textit{Asymmetric Coord} layouts from \cite{zsceval}, we found it difficult to train high-performing policies using the base VDN algorithm in a reasonable number of timesteps on these layouts, and therefore left them out in this work. Finally, we add a simple \textit{Large Room} layout in which agents have to travel long distances, encouraging specialization into more local workloads to avoid wasting time on movement.



\end{document}